\documentclass[twocolumn,abs,prb,showpacs]{revtex4-1}

\usepackage{graphicx}
\usepackage{dcolumn}
\usepackage{float}
\usepackage{amsmath}
\usepackage{amssymb}
\usepackage{bm}
\usepackage[mathscr]{euscript}

\newcommand{\comment}[1]{}

\newcommand{\wcr}{\omega_{\text{\tiny{CR}}}}

\begin{document}

%\preprint{}

\title{Cyclotron Resonance in Topological Insulators: Non-Relativistic Effects}

\author{C. J. Tabert$^{1,2}$}
\author{J. P. Carbotte$^{3,4}$}
\affiliation{$^1$Department of Physics, University of Guelph,
Guelph, Ontario N1G 2W1 Canada} 
\affiliation{$^2$Guelph-Waterloo Physics Institute, University of Guelph, Guelph, Ontario N1G 2W1 Canada}
\affiliation{$^3$Department of Physics, McMaster University,
Hamilton, Ontario L8S 4M1 Canada} 
\affiliation{$^4$Canadian Institute for Advanced Research, Toronto, Ontario M5G 1Z8 Canada}
\date{\today}

\begin{abstract}
{The low-energy Hamiltonian used to describe the dynamics of the helical Dirac fermions on the surface of a topological insulator contains a subdominant non-relativistic (Schr\"odinger) contribution.  This term can have an important effect on some properties while having no effect on others.  The Hall plateaus retain the same relativistic quantization as the pure Dirac case.  The height of the universal interband background conductivity is unaltered, but its onset is changed.  However, the non-relativistic term leads directly to particle-hole asymmetry.  It also splits the interband magneto-optical lines into doublets.  Here, we find that, while the shape of the semiclassical cyclotron resonance line is unaltered, the cyclotron frequency and its optical spectral weight are changed.  There are significant differences in both of these quantities for a fixed value of chemical potential or fixed doping away from charge neutrality depending on whether the Fermi energy lies in the valence or conduction band.
}
\end{abstract}

%Magneto-optical effects, Optical properties of low-D systems, LLs, Electronic transport in graphene

\maketitle
% body of paper here

\section{Introduction}

Recently, the metallic states that exist at the surface of a topological insulator (TI) have been extensively studied\cite{Kane:2005,Hasan:2010,Qi:2011,Moore:2010,Moore:2007,Fu:2007,Hsieh:2008}.  These display novel physical properties which could possibly be exploited in device applications such as quantum computing\cite{Fu:2008}.  This has lead to a flurry of activity both experimentally and theoretically which has provided new understanding.  Many new materials have been discovered\cite{Zhang:2009,Ando:2013}; however, there remains a need to increase the bulk band gap so as to more clearly separate the metallic surface states from the bulk insulating spectrum.  The surface states possess an odd number of Dirac points.  Here, for simplicity, it is sufficient to consider a single Dirac cone centred at the $\Gamma$ point of the surface Brillouin zone as in Bi$_2$Se$_3$\cite{Hsieh:2009}.  The Hamiltonian which governs the dynamics of the surface fermions involves real spin\cite{Hsieh:2009a} in contrast to graphene and other related materials where instead pseudospin is involved\cite{Geim:2007,Neto:2009,ZLi:2012,Stille:2012,Tabert:2013a}.  This is due to a strong spin-orbit coupling which provides a dominant linear-in-momentum Dirac contribution to the Hamiltonian with Fermi velocity $v_F$.  There is also a smaller non-relativistic contribution\cite{Liu:2010}, which is quadratic in momentum and characterized by a Schr{\"o}dinger mass $m$.  In this paper, we are particularly interested in understanding the changes in the magneto-optical properties of the helical surface fermions that result from the introduction of the mass term\cite{ZLi:2013}.

The Schr\"odinger mass term is a prominent feature of the low-energy continuum $\bm{k}\cdot\bm{p}$ Hamiltonians\cite{Liu:2010,Zhang:2009} which form the basis of our understanding of the TI surface band structure. Indeed, such work was critical in the discovery of three-dimensional (3D) TIs.  These simplified Hamiltonians are based on density-functional electronic-structure calculations, as first discussed by Zhang \emph{et al.}\cite{Zhang:2009} and elaborated upon in Ref.~\cite{Liu:2010}.  For the specific cases of Bi$_2$Te$_3$ and Bi$_2$Se$_3$, the band masses are $m=0.16m_e$ and $0.09m_e$, respectively, where $m_e$ is the bare electron mass.  Thus, for a magnetic field of 1 Tesla, the respective Schr\"odinger magnetic energy scales are $\sim 1.25$ meV and $0.7$ meV\cite{ZLi:2013}.  The corresponding Dirac energy scale (due to the linear-in-momentum relativistic part of the Hamiltonian) is of order 10 meV (a factor of ten larger than the non-relativistic contribution).  The two scales become comparable when the magnetic field $B$ is 40T.  This results from the Schr\"odinger terms linear-in-$B$ dependence compared to the $\sqrt{B}$ scaling of the relativistic piece.  Therefore, it is important to retain the quadratic-in-momentum part of the Hamiltonian when describing the surface states.  It has been included in the extensive literature\cite{Zhang:2009,Fuchs:2010,Wright:2013,Goerbig:2014,Tabert:2015,Tabert:2015b,Taskin:2011a,Ando:2013} (both theoretical and experimental) that exists on the phase offset seen in quantum oscillations of the Shubnikov-de Haas and de Haas-van Alphen effects.  There are numerous other optical studies in TIs including many in the infrared\cite{LaForge:2010,Xi:2013,Chapler:2014}; the most relevant for this work are those in the THz range\cite{Sushkov:2010,Jenkins:2013}.  Here, we mention specifically that of Wu \emph{et al.}\cite{Wu:2015} which provides information on the cyclotron frequency itself and, consequently, directly impacts the present work.  After the subtraction of a phonon and a lattice polarization term, the cyclotron resonance peak is extracted.  It was found that the quadratic-in-momentum term is essential in obtaining a consistent fit to this cyclotron resonance data\cite{Wu:2015}.  In another study of cyclotron resonance data, the Schr\"odinger piece\cite{Hancock:2011} was found to be dominant in strained films of HgTe (an intrinsic TI with negligible bulk carriers\cite{Brune:2011}).

Initially studied TIs where not truly insulating in the bulk and the topologically protected surface states could, thus, be affected by undesirable mixing with the bulk charge carriers.  Many strategies have been used to separate out the surface fermions.  In the magneto-optical study of Bi$_{0.91}$Sb$_{0.09}$ by Schafgans \emph{et al.}\cite{Schafgans:2012}, several Landau level (LL) series are observed.  They all display energy dependences that vary as the square root of the applied magnetic field and none are found to originate from the bulk bands since they do not extrapolate to the band gap energy.  In Bi$_{1-x}$Sb$_x$, the observed quantum oscillations correspond to a two-dimensional Fermi surface and hence, are identified with the surface states\cite{Schafgans:2012,Taskin:2009}.  They also observe a contribution form a 3D Fermi surface which they assign to coupling between the bulk and surface states.

It has also been possible to separate the surface from the bulk using time-resolved techniques.  The surface and bulk states' transient THz optical response has been found to correspond to very different time scales\cite{Valdes:2015}.  Crepaldi \emph{et al.}\cite{Crepaldi:2013} used time-resolved angular-resolved photoemission spectroscopy (ARPES) to differentiate between the two types of states. Such techniques have recently been used\cite{Sobota:2014} to show that the $A_{1g}$ phonon mode in Bi$_2$Se$_3$ is coupled to both the bulk and surface fermions and that this phonon softens on the surface.  There are many works on the electron-phonon interaction in TIs (see, for example, Refs.~\cite{Hatch:2011,Pan:2012,Kondo:2013}). High resolution ARPES is a particularly useful technique for this kind of investigation.  Kinks in the dispersion curves of the charge carriers are interpreted as coupling to possible bosons.  Some experiments\cite{Kondo:2013} have given large values for the coupling while others\cite{Hatch:2011} give much smaller values.  In one study, the coupling was found to be even smaller and interpreted to imply the topological surface states are not only protected from elastic scattering off impurities but also from scattering off low-energy phonons\cite{Pan:2012}.  Such effects are beyond the scope of this paper.

Work has also been done to entirely suppress the metallic character of the bulk states.  For example, this can be achieved by doping Cu in Bi$_2$Se$_3$\cite{Brahlek:2014,Lawson:2014}.  One can then study surface fermions without bulk influences.  This is also the case in Ti$_{1-x}$Bi$_{1+x}$Se$_2$ as documented in the work of Eguchi \emph{et al.}\cite{Eguchi:2014} and Kuroda \emph{et al.}\cite{Kuroda:2010}.

In the pure relativistic case, there is particle-hole symmetry between the conduction and valence bands which are both described by the same perfect Dirac cone.  The introduction of a Schr{\"o}dinger mass distorts these cones.  The conduction band narrows with increasing energy while the valence band fans out as the energy is lowered.  This yields an hourglass shape to the overall band structure.  The band bending leads to particle-hole asymmetry.  In this case, doping with holes rather than electrons will be different than the pure relativistic case.  Here, we will emphasize these differences.

When a magnetic field is applied perpendicular to the surface of the TI, Landau levels form.  In the pure relativistic system, their energies are characterized by the single Dirac magnetic energy scale $E_1=\hbar v_F\sqrt{eB/\hbar}$ where $e$ is the elementary charge.  When we account for the subdominant non-relativistic contribution, a second Schr{\"o}dinger magnetic energy scale $E_0=\hbar eB/m$ enters and modifies the eigenenergies of the LLs.  In particular, the level at zero energy which is so important in the relativistic case is moved to positive energy ($E_0/2$).  While for small values of $B$, $E_1$ is always larger than $E_0$. These two scales become comparable in size as $B$ is increased.  This leads to several known effects such as the splitting of the interband absorption lines in the longitudinal conductivity\cite{ZLi:2013,Tabert:2015b}.  At the same time, the Hall plateaus in the transverse DC conductivity retain\cite{ZLi:2014,Tabert:2015,Tabert:2015b} their relativistic quantization which is distinct from the non-relativistic values which arise when $\hbar eB/m$ is the dominant scale\cite{ZLi:2014}.

Even when $B$ is small and the semiclassical regime is considered, for which the chemical potential $|\mu|$ is much greater than both $E_1$ and $E_0$, important modifications to the cyclotron resonance frequency ($\omega_{\text{\tiny{CR}}}$) and its optical spectral weight ($W_{\rm CR}$) arise when $E_0$ is introduced.  Here we provide a simple analytic formula for the resulting line shape and study the particle-hole asymmetry in both $\omega_{\text{\tiny{CR}}}$ and $W_{\rm CR}$.  While it is the intraband optical transitions which are central to the semiclassical regime, the evolution of the interband transitions in the same limit of $|\mu|\gg E_1$ and $E_0$ are also of interest.  The magnitude of the universal background which emerges is completely independent of the Schr{\"o}dinger mass but its onset is not\cite{ZLi:2015}.  In particular, in this work we emphasize the particle-hole asymmetry of the onset.

The paper is organized as follows: In Sec.~II, we introduce the Kubo formula for both the longitudinal and transverse (Hall) magneto-optical conductivity\cite{ZLi:2013,Tabert:2015b}.  We also present results for both Re$\sigma_{xx}(\Omega)$ and Re$\sigma_{xy}(\Omega)$ at several values of $B$ and feature their evolution as $B$ is decreased sufficiently that the LL structures become smooth and reveal a background.  In Sec.~III, we start from the finite $B$ formulas for the real part of the longitudinal conductivity and show how the semiclassical regime is obtained.  The end result is analytic formulae for the cyclotron resonance lineshape, frequency and optical spectral weight.  We retain the appropriate limit as $E_0\rightarrow 0$.  We study the approach to the semiclassical limit and discuss the particle-hole asymmetry that arises.  We compare both a constant $|\mu|$ for $\mu\lessgtr 0$ and constant doping level $n$ for electrons and holes.  In Sec.~IV, we provide a discussion of the universal interband background\cite{ZLi:2015} and its particle-hole asymmetry.  An alternative derivation, based on the finite $B$ formula for the conductivity, is provided for the interband background which is additional to the intraband cyclotron resonance line.  Details about the asymmetry of the absorption process onsets are provided.  Finally, in Sec.~V, we give a discussion of our new results and draw conclusions.

\section{Formalism}

\begin{figure}[h!]
\begin{center}
\includegraphics[width=1.0\linewidth]{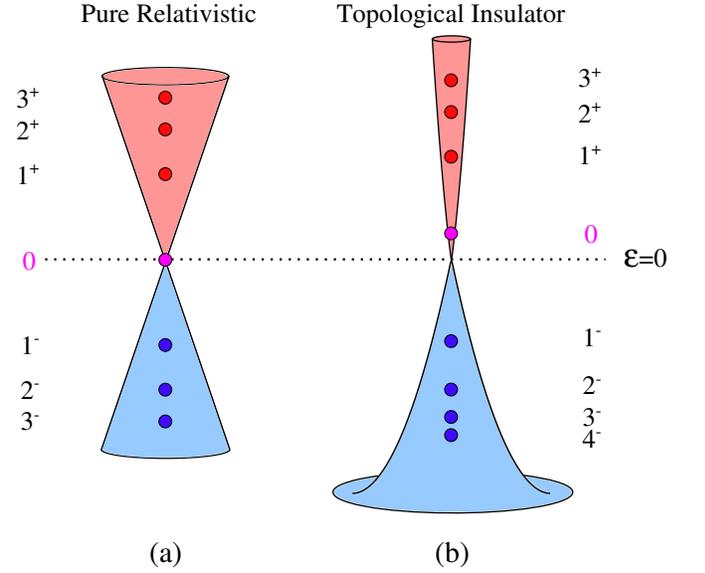}
\end{center}
\caption{\label{fig:LLs}(Color online) (a) Perfect Dirac surface states in the pure relativistic limit. (b) Hourglass shaped spectrum when a subdominant non-relativistic term is present.  The solid dots with labels $0,1^\pm, 2^\pm, ...$ show the energy of the LLs in a magnetic field.  The shading indicates the conduction (red) and valence (blue) bands.
}
\end{figure}

The model low-energy Hamiltonian on which this work is based includes a small non-relativistic Schr{\"o}dinger term characterized by a mass $m$ and proportional to the square of the momentum.  In addition, we include a dominant relativistic linear-in-momentum Dirac term with Fermi velocity $v_F$.  It has the form\cite{Bychkov:1984, Bychkov:1984a}
\begin{align}\label{HAM}
\hat{H}=\frac{\hbar^2k^2}{2m}+\hbar v_F(k_x\hat{\sigma}_y-k_y\hat{\sigma}_x),
\end{align} 
where $\hat{\sigma}_x$ and $\hat{\sigma}_y$ are the Pauli spin matrices and $\bm{k}$ is the momentum relative to the $\Gamma$ point of the surface state Brillouin zone.  As previously discussed, representative values for the mass and velocity are $m\approx 0.1m_e$ (with $m_e$ the bare electron mass) and $v_F$ of order 10$^5$m/s, respectively\cite{Liu:2010}.  When a magnetic field $B$ is oriented perpendicular to the 2D surface states, the magnetic energy scales for the relativistic ($E_1$) and non-relativistic ($E_0$) pieces of the Hamiltonian are of order 10 meV and 1 meV, respectively, for $B=1$T.  By definition, $E_1=\hbar v_F\sqrt{eB/\hbar}$ and $E_0=\hbar eB/m$ both involve the magnetic length $l_B=\sqrt{\hbar/(eB)}$.  In the Landau gauge [with magnetic vector potential $\bm{A}=(0,Bx,0)$], the Hamiltonian can be written as
\begin{align}\label{HAM-Matrix}
H&=\left(\begin{array}{cc}
\displaystyle \frac{\hbar^2}{ml_B^2}\left[a^\dagger a+\frac{1}{2}\right] & \displaystyle -\hbar v_F\frac{\sqrt{2}}{l_B} a\\
\displaystyle -\hbar v_F\frac{\sqrt{2}}{l_B} a^\dagger & \displaystyle \frac{\hbar^2}{ml_B^2}\left[a^\dagger a+\frac{1}{2}\right]
\end{array}\right),
\end{align}  
where
\begin{align}
a^\dagger\equiv\frac{l_B}{\sqrt{2}}\left[-ik_x+\frac{x+k_yl_B^2}{l_B^2}\right],
\end{align}
and
\begin{align}
a\equiv\frac{l_B}{\sqrt{2}}\left[ik_x+\frac{x+k_yl_B^2}{l_B^2}\right],
\end{align}
are the creation and annihilation operators of the harmonic oscillator Hamiltonian.  The energy of the LLs is
\begin{align}\label{LL-TI}
\mathcal{E}_{N,s}=\left\lbrace\begin{array}{cc}
\displaystyle E_0N+s\sqrt{2NE_1^2+\frac{E_0^2}{4}} & N=1,2,3,...\\
\displaystyle\frac{E_0}{2} & N=0
\end{array}\right.,
\end{align}
where $s=\pm$ for the conduction and valence band, respectively. A schematic comparison between the LLs of the pure relativistic limit ($E_0=0$) and TI is shown in Fig.~\ref{fig:LLs}.
The LLs (dots) are overlaid on the respective $B=0$ band structures.  In the TI, the $N=0$ level is no longer pinned at zero energy but has shifted upwards.  The symmetry between the conduction (red) and valence (blue) levels is also lost with finite $E_0$.  In the following numerical work, we will take $E_1/\sqrt{B}=10.4$ meV$/\sqrt{\rm T}$ and $E_0/B=1.1$ meV/T as characteristic values of a TI.  In terms of the spin-up and spin-down amplitudes, the eigenfunctions of the state $\left|Ns\right\rangle$ are
\begin{align}
\left|Ns\right\rangle=\left(\begin{array}{c}
\mathcal{C}^\uparrow_{N,s}\left|N-1\right\rangle\\
\mathcal{C}^\downarrow_{N,s}\left|N\right\rangle
\end{array}\right),
\end{align}
where $a\left|N\right\rangle=\sqrt{N}\left|N-1\right\rangle$ and $a^\dagger\left|N\right\rangle=\sqrt{N+1}\left|N+1\right\rangle$.
The spin-up amplitude is
\begin{align}\label{Cup}
\mathcal{C}^\uparrow_{N,s}=\left\lbrace\begin{array}{cc}
\displaystyle -s\sqrt{\frac{1}{2}-s\frac{E_0/2}{2(\mathcal{E}_{N,+}-E_0N)}} & N=1,2,3,...\\
0 & N=0
\end{array}\right.,
\end{align}
and
\begin{align}\label{Cdown}
\mathcal{C}^\downarrow_{N,s}=\left\lbrace\begin{array}{cc}
\displaystyle\sqrt{\frac{1}{2}+s\frac{E_0/2}{2(\mathcal{E}_{N,+}-E_0N)}} & N=1,2,3,...\\
1 & N=0
\end{array}\right.,
\end{align}
gives the spin-down component.    

Using the Kubo formula, the dynamical conductivity $\sigma_{\alpha\beta}(\Omega)$ is given by
\begin{align}\label{Kubo-Mag}
\sigma_{\alpha\beta}(\Omega)=\frac{i\hbar}{2\pi l_B^2}&\sum_{\substack{N,M=0 \\s,s^\prime=\pm}}^\infty\frac{f_{M,s^\prime}-f_{N,s}}{\mathcal{E}_{N,s}-\mathcal{E}_{M,s^\prime}}\notag\\
&\times\frac{\left\langle \bar{N}s|\hat{j}_\alpha|\bar{M}s^\prime\right\rangle\left\langle \bar{M}s^\prime|\hat{j}_\beta|\bar{N}s\right\rangle}{\hbar\Omega+\mathcal{E}_{M,s^\prime}-\mathcal{E}_{N,s}+i\hbar/(2\tau)},
\end{align}
where $f_{N,s}$ is the Fermi function for state $N$ in band $s$, $\alpha$ and $\beta$ index the $x$ and $y$ components, $\tau$ is a phenomenological optical scattering time taken to be a large constant and $\hat{j}_{\alpha}=e\hat{v}_\alpha\equiv (e/\hbar)(\partial \hat{H}/\partial k_{\alpha})$ is the current operator.  At zero temperature, $f_{N,s}$ can be replaced by the Heaviside step function $\Theta(\mu-\mathcal{E}_{N,s})$; here, $\mu$ is the chemical potential.  The necessary velocity operators are
\begin{align}\label{vx}
\hat{v}_x=\frac{\hbar}{m}k_x+v_F\hat{\sigma}_y=i\frac{\hbar}{m}\frac{a^\dagger-a}{\sqrt{2}l_B}+v_F\hat{\sigma}_y,
\end{align}
and
\begin{align}\label{vy}
\hat{v}_y=\frac{\hbar}{m}\left[k_y+\frac{eBx}{\hbar}\right]-v_F\hat{\sigma}_x=\frac{\hbar}{m}\frac{a^\dagger+a}{\sqrt{2}l_B}-v_F\hat{\sigma}_x.
\end{align}
Calculating the appropriate matrix elements, we obtain the real part of the longitudinal conductivity\cite{Tabert:2015b}
\begin{align}\label{sigmaxx}
{\rm Re}\left\lbrace\frac{\sigma_{xx}(\Omega)}{e^2/\hbar}\right\rbrace &=\frac{E_1^2}{2\pi}\sum_{\substack{N,M=0 \\s,s^\prime=\pm}}^\infty\frac{f_{M,s^\prime}-f_{N,s}}{\mathcal{E}_{N,s}-\mathcal{E}_{M,s^\prime}}\\
&\times\frac{\eta}{(\Omega+\mathcal{E}_{M,s^\prime}-\mathcal{E}_{N,s})^2+\eta^2}\notag\\
&\times[\mathcal{F}(Ns;Ms^\prime)\delta_{N,M-1}+\mathcal{F}(Ms^\prime;Ns)\delta_{M,N-1}],\notag
\end{align}
where $\eta\equiv\hbar/(2\tau)$ is a constant phenomenological optical scattering rate and\cite{Tabert:2015b}
\begin{align}\label{FNM}
\mathcal{F}(Ns;Ms^\prime)&\equiv\left[\mathcal{C}^\uparrow_{M,s^\prime}\mathcal{C}^\downarrow_{N,s}-\frac{E_0}{\sqrt{2}E_1}\right.\\
&\times\left.\left(\sqrt{N}\mathcal{C}^\uparrow_{M,s^\prime}\mathcal{C}^\uparrow_{N,s}+\sqrt{N+1}\mathcal{C}^\downarrow_{M,s^\prime}\mathcal{C}^\downarrow_{N,s}\right)\right]^2.\notag
\end{align}
The imaginary part of the conductivity is found by replacing the $\eta$ factor in the numerator with $\Omega+\mathcal{E}_{M,s^\prime}-\mathcal{E}_{N,s}$. 
Similarly, the real part of the transverse Hall conductivity is\cite{Tabert:2015b}
\begin{align}\label{sigmaxy}
{\rm Re}\left\lbrace\frac{\sigma_{xy}(\Omega)}{e^2/\hbar}\right\rbrace &=\frac{E_1^2}{2\pi}\sum_{\substack{N,M=0 \\s,s^\prime=\pm}}^\infty\frac{f_{M,s^\prime}-f_{N,s}}{\mathcal{E}_{N,s}-\mathcal{E}_{M,s^\prime}}\\
&\times\frac{\Omega+\mathcal{E}_{M,s^\prime}-\mathcal{E}_{N,s}}{(\Omega+\mathcal{E}_{M,s^\prime}-\mathcal{E}_{N,s})^2+\eta^2}\notag\\
&\times[\mathcal{F}(Ns;Ms^\prime)\delta_{N,M-1}-\mathcal{F}(Ms^\prime; Ns)\delta_{M,N-1}].\notag
\end{align}
Here, the imaginary part is given by the replacement $\Omega+\mathcal{E}_{M,s^\prime}-\mathcal{E}_{N,s}\rightarrow -\eta$ in the numerator. 

\section{Numerical Results for the Dynamical Conductivity}

To set the stage for understanding the semiclassical limit, we start by examining how the dynamical conductivity evolves as the magnitude of the magnetic field $B$ is progressively decreased towards $B=0$.  As we are interested in modifications to the semiclassical limit brought about by the subdominant non-relativistic (Schr{\"o}dinger) contribution in the Hamiltonian [Eqn.~\eqref{HAM}], differences between the pure relativistic (Dirac) and TI are emphasized.
 
\begin{figure}[h!]
\begin{center}
\includegraphics[width=1.0\linewidth]{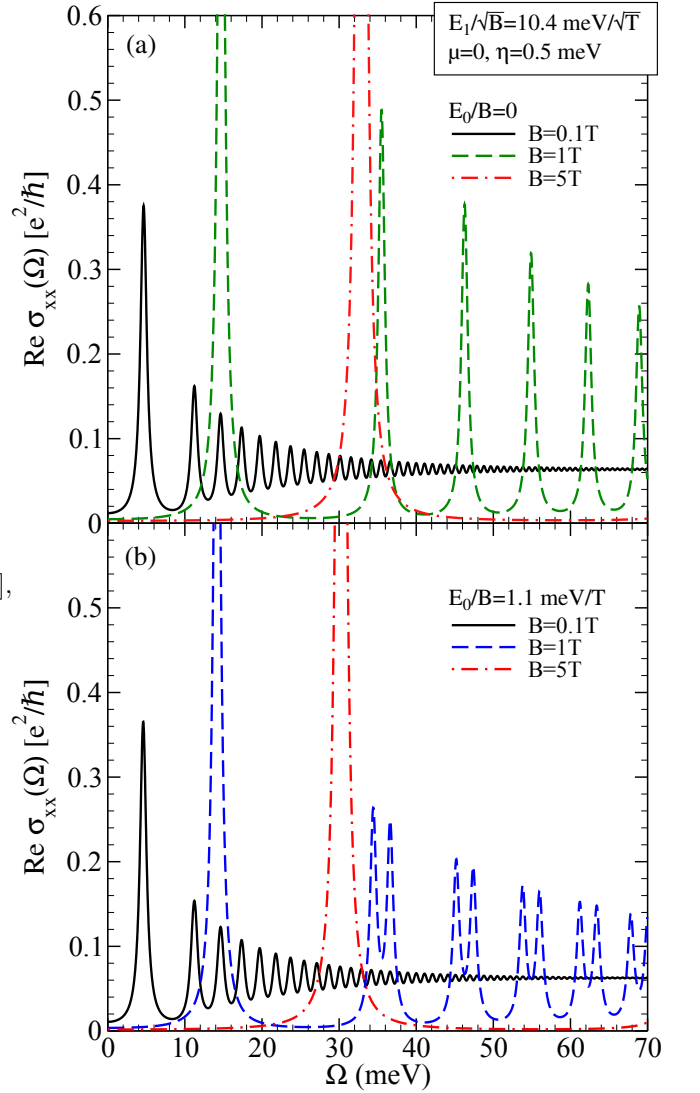}
\end{center}
\caption{\label{fig:Condxx-TI-Dirac}(Color online) Real part of the longitudinal optical conductivity as a function of photon energy at charge neutrality ($\mu=0$) for (a) the pure relativistic case and (b) with the addition of a non-relativistic contribution.  Three values of $B$ are considered: $B=0.1$T, $B=1$T and $B=5$T.
}
\end{figure}

In Fig.~\ref{fig:Condxx-TI-Dirac}, we compare Re$\sigma_{xx}(\Omega)$ for a typical TI with the ideal relativistic system.  
The top frame is for $E_0=0$ (i.e. no Schr{\"o}dinger mass term).  In both cases, we take the Dirac magnetic energy scale to be $E_1/\sqrt{B}=10.4$ meV$/\sqrt{\rm T}$ and set the chemical potential at charge neutrality ($\mu=0$).  Three values of magnetic field are considered: B=0.1T (solid black curve), B=1T [dashed green/blue curve for (a)/(b)] and B=5T (dash-dotted red curve).  Therefore, $E_1=3.29$ meV, 10.5 meV and 23.3 meV, respectively.  It is these energies that determine the energies of the various LLs.  For $E_0=0$, Eqn.~\eqref{LL-TI} gives $\mathcal{E}_{N,s}=s\sqrt{2NE_1}$ for $N=1,2,3,...$ and $\mathcal{E}_0=0$ for $N=0$.  In the pure relativistic case, one has perfect particle-hole symmetry about the neutrality point.  In the numerical work, we have taken the optical scattering rate $\eta$ of Eqn.~\eqref{sigmaxx} to be 0.5 meV which provides a width to the optical absorption lines that otherwise correspond to Dirac $\delta$-functions.  First, we note that a universal interband background becomes well defined at high photon frequencies in the solid black curve.  In units of $e^2/\hbar$, this takes the value of 1/16.  For larger magnetic fields (dashed blue and dash-dotted red curves), the energy spacing between the optical peaks has increased and the background is no longer resolved.  In all cases, the first line is an intraband transition from $N=0$ to $(N,s)=(1,+)$ while all other lines are due to interband transitions from $(N,-)\rightarrow (N\pm 1,+)$ with $N=1,2,3,...$.  The two interband lines for a given $N$ are degenerate in energy for a given value of $N$.  These results are for comparison with those presented in the lower frame which apply to the case when the Schr{\"o}dinger energy $E_0$ is not set at zero ($E_0/B=1.1$ meV/T).  For $B=0.1$T, $E_0$ is small compared to $E_1$.  For $B=5$T, it has increased to 5.5 meV which is roughly a factor of four smaller than $E_1$.  Still, the energies at which the various lines occur have not shifted a great deal when compared to the top frame; however, we can clearly see that the intraband line for $B=5$T has shifted downward upon the introduction of a Schr{\"o}dinger term.  As is clearly seen in the dashed blue curve, for $B=1$T, the interband transition peak has split into two with the energy of the splitting given by the value of $E_0$.  This can be traced to the fact that, in a TI, the lines $(N,-)\rightarrow (N+1,+)$ and $(N,-)\rightarrow(N-1,+)$ are no longer degenerate in energy.  As $B$ is reduced in magnitude, the splitting of the interband peaks into doublets is no longer resolved (see the solid black curve for $B=0.1$T).  This is because $E_0$ itself becomes small and the residual scattering rate that we have used ($\eta=0.5$ meV) washes out the splitting.  The universal background seen at higher values of photon energy $\Omega$ rapidly becomes constant and equal to $e^2/(16\hbar)$, completely independent of the Schr{\"o}dinger mass term.  In fact, because we have taken $\mu=0$ (charge neutrality), the entire curve becomes constant independent of $\Omega$ for $B=0$.  

\begin{figure}[]
\begin{center}
\includegraphics[width=1.0\linewidth]{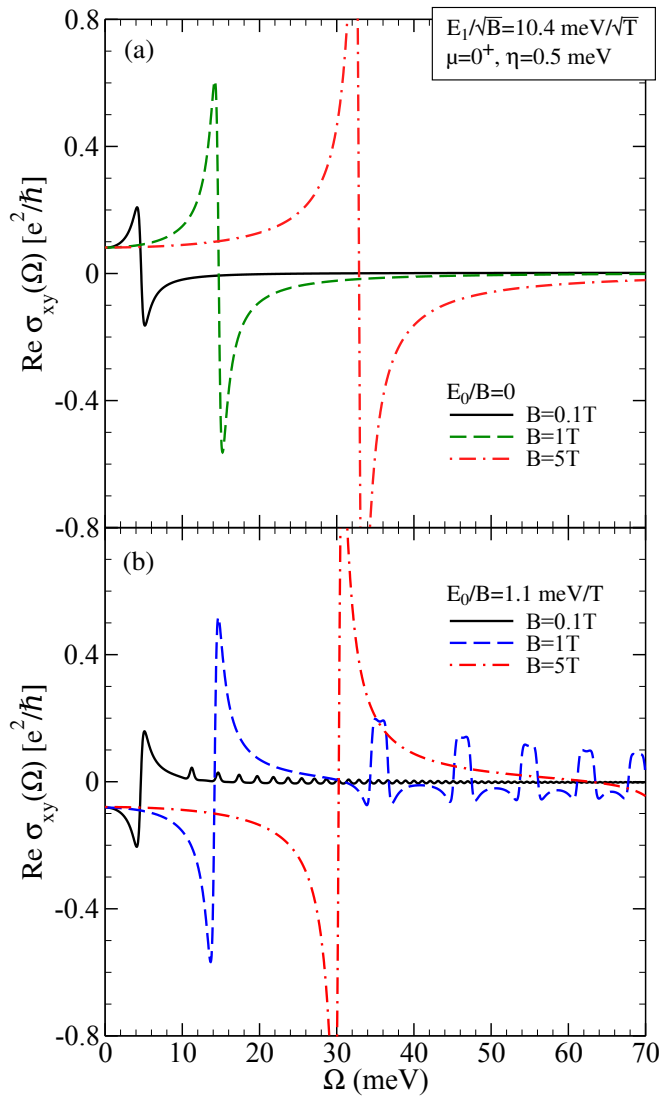}
\end{center}
\caption{\label{fig:Condxy-TI-Dirac}(Color online) Real part of the transverse Hall conductivity for the same parameters as Fig.~\ref{fig:Condxx-TI-Dirac}.
}
\end{figure}

It is illuminating to also consider the evolution of the Hall conductivity as $B\rightarrow 0$.  We know that at $B=0$, this quantity [Re$\sigma_{xy}(\Omega)$] must be zero rather than the constant value seen above for Re$\sigma_{xx}(\Omega)$.  In Fig.~\ref{fig:Condxy-TI-Dirac}, we provide a comparison between the pure Dirac and TI for the real part of the Hall conductivity.
The same three values of magnetic field are considered and the notation is the same as for Fig.~\ref{fig:Condxx-TI-Dirac}.  Again, the energy of the first peak/valley is only slightly affected by the introduction of an $E_0$ term.  However, there are important differences which must be noted.  Namely, in the top frame, the peak is always first followed by a valley at slightly higher energy.  In the bottom frame, it is the opposite.  Associated with this reversal is the fact that the DC limit of Re$\sigma_{xy}(\Omega)$ which gives the quantized Hall conductivity has the same magnitude ($1/(4\pi)$ in our units) but is of opposite sign.  This change has been brought about by the introduction of the Schr{\"o}dinger term $E_0$; this does not change the quantization in the plateau height (which remains relativistic) but does change the energy (chemical potential) at which the step to a new plateau occurs.  For a TI, the step from $-1/2$ to $1/2$ quantization (in units of $e^2/h$) occurs at finite $\mu$ rather than at $\mu=0^+$.  This is due to the first LL sitting at an energy $E_0/2$ rather than at zero.  Some further explanation of our language is required.  As we have stated, in our units, the universal interband background has a value of $1/16$ or Re$\sigma_{xx}(\Omega)=e^2/(16\hbar)=e^2\pi/(8h)$ which is a factor of four smaller than the well known value of $e^2\pi/(2h)$ for graphene.  In graphene, this value accounts for a valley-spin degeneracy which is not present in a TI.  Further, the Hall plateau quantization is $\pm 1/2, \pm 3/2, \pm 5/2,...$ in graphene (in units of $e^2/h$) if the four fold degeneracy is removed.  Another striking difference between the pure relativistic [frame (a)] and the TI case [frame (b)] is the additional LL structure seen in the lower frame which is absent in the upper frame.  This is traced to the splitting of the optical transitions between $N-1\rightarrow N$ and $N\rightarrow N+1$ for a TI.  These oscillations are clearly significant at larger values of $B$ but are progressively reduced as $B$ becomes small.  Note that for the $B=0.1$T (solid black curve), Re$\sigma_{xy}(\Omega)\sim 0$ for most of the $\Omega$ range shown.  At $B=0$, the entire curve is of course zero.

\begin{figure}[h!]
\begin{center}
\includegraphics[width=1.0\linewidth]{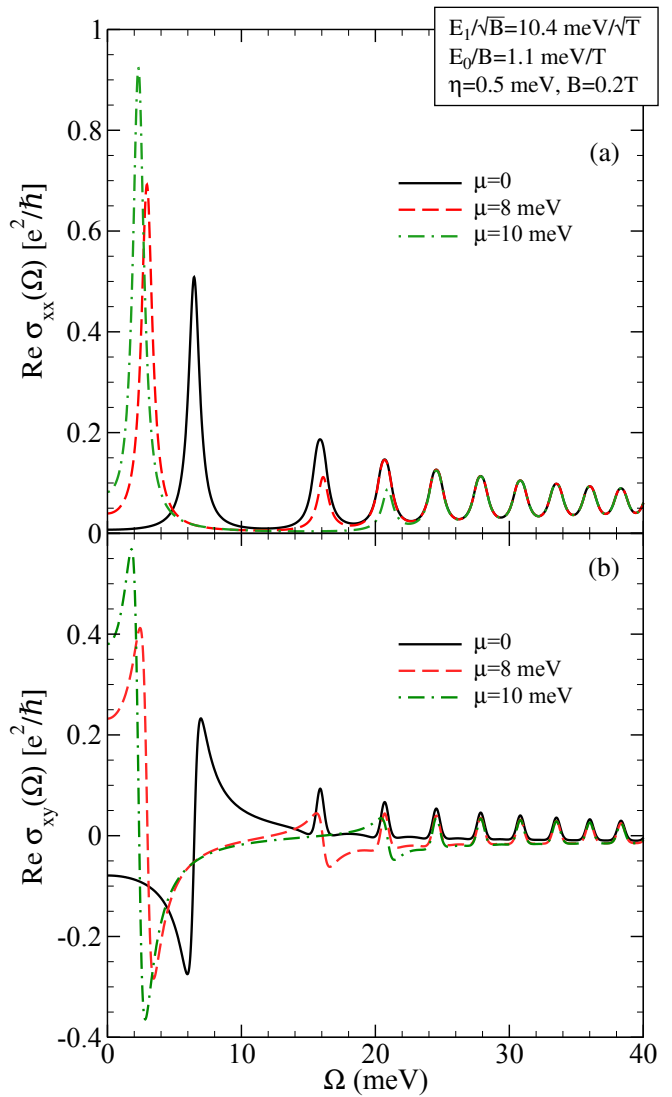}
\end{center}
\caption{\label{fig:Condxx-xy}(Color online) (a) Real part of the longitudinal AC conductivity as a function of photon energy for $B=0.2$T and three values of chemical potential: $\mu=0$ (charge neutrality), $\mu=8$ meV and $\mu=10$ meV. (b) Real part of the Hall conductivity for the same parameters as (a).
}
\end{figure}

So far, we have considered $\mu=0$ to get a first understanding of the differences and unchanged characteristics of the magneto-optics which are brought about by the introduction of a subdominant Schr{\"o}dinger piece in the Hamiltonian.  Next, we will concentrate on finite $\mu$.  In fact, our interest will be on the semiclassical regime for $\mu$ much larger than the magnetic scales $E_1$ and $E_0$.  In Fig.~\ref{fig:Condxx-xy}, we show the results for the real parts of the longitudinal conductivity Re$\sigma_{xx}(\Omega)$ vs. $\Omega$ [frame (a)] and the Hall conductivity Re$\sigma_{xy}(\Omega)$  vs. $\Omega$ [frame (b)].  
The magnetic field is set at $B=0.2$T and three values of $\mu$ are considered.  The solid black curve corresponds to $\mu=0$ and is for comparison.  The dashed red curve applies to $\mu=8$ meV and the dash-dotted green line is for $\mu=10$ meV.  We want to focus on the first intraband peak and understand its evolution as $\mu$ is increased.  It is this line which eventually evolves into the cyclotron resonance line of the semiclassical theory.  We note that, as $\mu$ increases, the position of this peak moves to lower energies and acquires added intensity (optical spectral weight).  Frame (b) gives Re$\sigma_{xy}(\Omega)$.  We note that, in these curves, the position in energy of the first valley-to-peak (for $\mu=0$) and peak-to-valley (for finite $\mu$) structures also evolves towards lower energies and tracks the intraband peaks in Re$\sigma_{xx}(\Omega)$ of the top frame.  In particular, we emphasize the DC value of Re$\sigma_{xy}(\Omega)$ which gives the Hall plateaus.  For $\mu=0$, it is $-1/2$ [i.e. $-e^2/(4\pi\hbar)$] while for $\mu=8$ meV it has moved to $3/2$ and then to $5/2$ (in units of $e^2/h$) for $\mu=10$ meV.   These values are independent of the mass term in the Hamiltonian.  We now turn to the semiclassical regime $\mu\gg E_1$ and $E_0$.

\section{Semiclassical Limit}

\begin{figure}[h!]
\begin{center}
\includegraphics[width=1.0\linewidth]{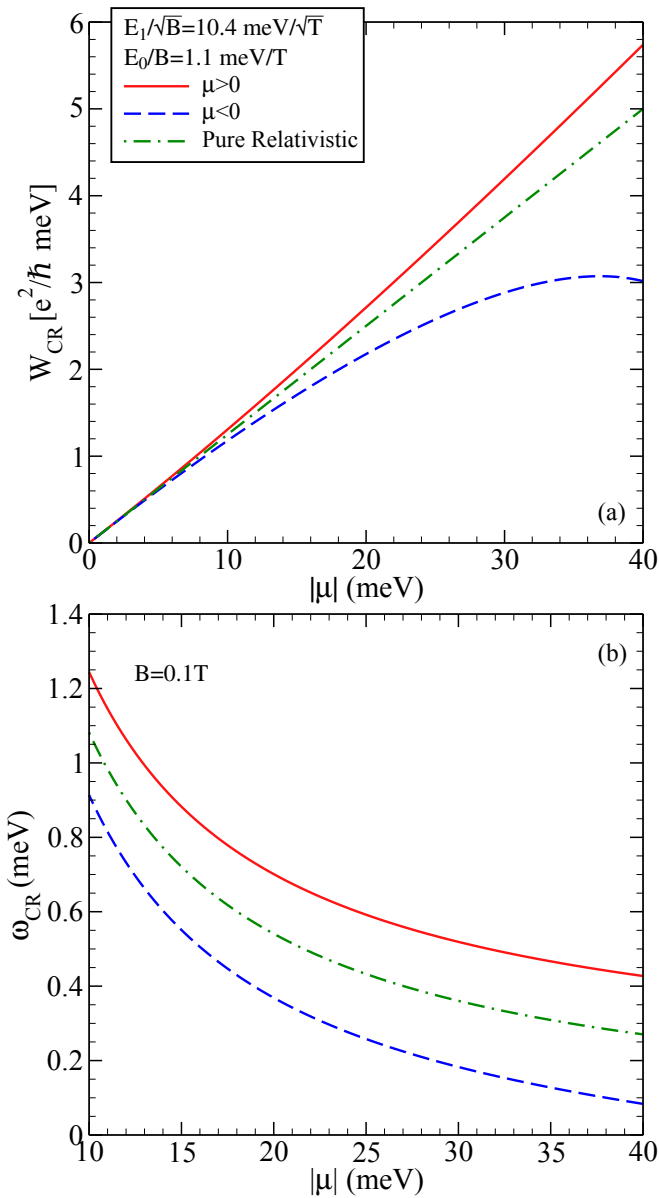}
\end{center}
\caption{\label{fig:Cyc}(Color online) (a) Optical spectral weight under the cyclotron resonance line as a function of $|\mu|$ for the pure relativistic and TI regimes.  (b) The corresponding results for the cyclotron frequency at $B=0.1$T.
}
\end{figure}

We begin with Eqn.~\eqref{sigmaxx} for the real part of the AC longitudinal conductivity.  We work in the limit of $\mu$ much larger than the magnetic quantization such that the chemical potential falls between the $N_c^{\rm th}$ and $(N_c+1)^{\rm th}$ LLs with $N_c$ very large and given by
\begin{align}\label{Nc}
\mu=E_0N_c+\sqrt{2N_cE_1^2+\left(\frac{E_0}{2}\right)^2}.
\end{align} 
For definiteness, we assume $\mu>0$ (i.e. the conduction band is partially filled).  In this case
\begin{align}
{\rm Re}\left\lbrace\frac{\sigma_{xx}(\Omega)}{e^2/\hbar}\right\rbrace &=\frac{E_1^2}{2\pi}\sum_{N=0}^\infty \frac{f_{N,+}-f_{N-1,+}}{\mathcal{E}_{N-1,+}-\mathcal{E}_{N,+}}\\
&\times\left\lbrace\frac{\eta\mathcal{F}(N-1+;N+)}{(\Omega+\mathcal{E}_{N,+}-\mathcal{E}_{N-1,+})^2+\eta^2}\right.\notag\\
&\left.+\frac{\eta\mathcal{F}(N-1+;N+)}{(\Omega+\mathcal{E}_{N-1,+}-\mathcal{E}_{N,+})^2+\eta^2}\right\rbrace.\notag
\end{align}
From here on, the $+$ index shall be implied (and suppressed) as we are only interested in $s,s^\prime=+$.  In our limit, the first factor of the sum becomes $-\partial f_N/\partial \mathcal{E}_N$ and the energy difference $\mathcal{E}_N-\mathcal{E}_{N-1}\equiv\omega_{\text{\tiny{CR}}}(\mu)$ is the cyclotron resonance energy for $\mu>0$.  At zero temperature,
\begin{align}\label{sigxx-sum}
{\rm Re} &\left\lbrace\frac{\sigma_{xx}(\Omega)}{e^2/\hbar}\right\rbrace =\frac{E_1^2}{2\pi}\sum_{N=0}^\infty \delta\left(\mu-\mathcal{E}_N\right)\mathcal{F}(N-1;N)\\
&\times\left\lbrace\frac{\eta}{(\Omega+\omega_{\text{\tiny{CR}}})^2+\eta^2}+\frac{\eta}{(\Omega-\omega_{\text{\tiny{CR}}})^2+\eta^2}\right\rbrace.\notag
\end{align}
For $B\rightarrow 0$, $E_0\ll E_1$ and Eqn.~\eqref{Nc} gives
\begin{align}\label{Nc-approx}
\sqrt{2N_c}=\frac{E_1}{E_0}\left\lbrace\sqrt{1+\frac{2\mu }{mv_F^2}}-1\right\rbrace.
\end{align}
The sum over $N$ in Eqn.~\eqref{sigxx-sum} also changes into an integral.  Therefore, the Dirac $\delta$-function $\delta(\mu-\mathcal{E}_N)$ means that the entire integrand is to be evaluated at $N=N_c$.  This gives
\begin{align}
{\rm Re} &\left\lbrace\frac{\sigma_{xx}(\Omega)}{e^2/\hbar}\right\rbrace =\frac{E_1^2}{2\pi}\frac{1}{\partial\mathcal{E}_N/\partial N\bigg|_{N_c}}\mathcal{F}(N_c-1;N_c)\\
&\times\left\lbrace\frac{\eta}{(\Omega+\omega_{\text{\tiny{CR}}})^2+\eta^2}+\frac{\eta}{(\Omega-\omega_{\text{\tiny{CR}}})^2+\eta^2}\right\rbrace.\notag
\end{align}
But
\begin{align}
\left.\frac{\partial\mathcal{E}_N}{\partial N}\right|_{N_c}=\frac{E_0\sqrt{1+\frac{2\mu}{mv_F^2}}}{\sqrt{1+\frac{2\mu }{mv_F^2}}-1},
\end{align}
and hence
\begin{align}\label{sigxx-cyc}
{\rm Re} &\left\lbrace\frac{\sigma_{xx}(\Omega)}{e^2/\hbar}\right\rbrace =\frac{mv_F^2}{2}\frac{\sqrt{1+\frac{2\mu }{mv_F^2}}-1}{\sqrt{1+\frac{2\mu}{mv_F^2}}}\mathcal{F}(N_c-1;N_c)\\
&\times\left\lbrace\frac{\eta/\pi}{(\Omega+\omega_{\text{\tiny{CR}}})^2+\eta^2}+\frac{\eta/\pi}{(\Omega-\omega_{\text{\tiny{CR}}})^2+\eta^2}\right\rbrace.\notag
\end{align}
The only $\Omega$ dependence that remains is in the curly bracket and the integral over photon energy of this term is normalized to a value of one.  It provides an analytic formula for the frequency profile of the cyclotron resonance line.  The remaining factors in Eqn.~\eqref{sigxx-cyc} give the optical spectral weight of the cyclotron resonance which we denote $W_{CR}$, where
\begin{align}\label{WCR}
W_{CR}=\frac{e^2}{\hbar}\frac{mv_F^2}{2}\frac{\sqrt{1+\frac{2\mu }{mv_F^2}}-1}{\sqrt{1+\frac{2\mu}{mv_F^2}}}\mathcal{F}(N_c-1;N_c).
\end{align}
We now need to evaluate $\mathcal{F}(N_c-1;N_c)$.  Since $N_c$ is very large as $B\rightarrow 0$, we can replace $N_c-1$ by $N_c$ to get
\begin{align}
\mathcal{F}(N_c;N_c)&\approx\left[\mathcal{C}^\uparrow_{N_c}\mathcal{C}^\downarrow_{N_c}-\frac{E_0}{\sqrt{2}E_1}\right.\\
&\times\left.\left(\sqrt{N_c}\mathcal{C}^\uparrow_{N_c}\mathcal{C}^\uparrow_{N_c}+\sqrt{N_c}\mathcal{C}^\downarrow_{N_c}\mathcal{C}^\downarrow_{N_c}\right)\right]^2,\notag
\end{align}
which follows from Eqn.~\eqref{FNM}.  Using Eqns.~\eqref{Cup} and \eqref{Cdown}, we note that
\begin{align}
\mathcal{C}^\uparrow_{N_c}\approx -\sqrt{\frac{1}{2}-\frac{E_0/2}{2\sqrt{2N_c}E_1}}
\end{align}
and
\begin{align}
\mathcal{C}^\downarrow_{N_c}\approx \sqrt{\frac{1}{2}+\frac{E_0/2}{2\sqrt{2N_c}E_1}}.
\end{align}
Using Eqn.~\eqref{Nc-approx} for $\sqrt{2N_c}$, we find
\begin{align}
\frac{E_0}{\sqrt{2N_c}E_1}=\frac{E_0}{mv_F^2}\frac{1}{\sqrt{1+\frac{2\mu}{mv_F^2}}-1}
\end{align}
which goes to zero as $B\rightarrow 0$.  Therefore, $\mathcal{C}^\uparrow_{N_c}\approx-\sqrt{1/2}$ and $\mathcal{C}^\downarrow_{N_c}\approx\sqrt{1/2}$.  Thus
\begin{align}
\mathcal{F}(N_c;N_c)&\approx\left(-\frac{1}{2}-\frac{E_0}{\sqrt{2}E_1}\sqrt{N_c}\right)^2=\frac{1}{4}\left(1+\frac{2\mu }{mv_F^2}\right).
\end{align}
When this is substituted into Eqn.~\eqref{WCR}, we obtain
\begin{align}
W_{CR}=\frac{e^2}{\hbar}\frac{mv_F^2}{8}\left[\sqrt{1+\frac{2\mu}{mv_F^2}}-1\right]\sqrt{1+\frac{2\mu }{mv_F^2}},
\end{align}
for the optical spectral weight of the cyclotron resonance line.  Including negative chemical potentials, we find
\begin{align}
W_{CR}=\frac{e^2}{\hbar}\frac{mv_F^2}{8}\left|\sqrt{1+\frac{2\mu}{mv_F^2}}-1\right|\sqrt{1+\frac{2\mu }{mv_F^2}},
\end{align}
for $\mu \lessgtr 0$.  The profile in photon energy of the line is given by
\begin{align}
\frac{\eta/\pi}{(\Omega+\omega_{\text{\tiny{CR}}})^2+\eta^2}+\frac{\eta/\pi}{(\Omega-\omega_{\text{\tiny{CR}}})^2+\eta^2}
\end{align}
for both $\mu \lessgtr 0$.  This formula is unchanged from the result found for graphene\cite{Gusynin:2009} except that the cyclotron frequency $\wcr$ has been altered.  By definition, the cyclotron resonance energy $\wcr$ at chemical potential $\mu$ is given by $\mathcal{E}_{N_c}-\mathcal{E}_{N_c-1}\equiv\wcr$ in the limit of large $N_c$ determined by $\mu$.  We get
\begin{align}\label{wcr}
\wcr(\mu)=\frac{\hbar eB v_F^2}{2|\mu|}\left(1+\sqrt{1+\frac{2\mu}{mv_F^2}} \right)\sqrt{1+\frac{2\mu }{mv_F^2}},
\end{align}
which does depend on the Schr\"odinger mass.  In Fig.~\ref{fig:Cyc}(a), we plot the optical spectral weight ($W_{CR}$) of the cyclotron resonance line in a TI as a function of $|\mu|$.  
Three cases are considered. The dash-dotted green curve is for comparison and corresponds to the pure relativistic limit.  In this instance, $W_{CR}$ increases linearly with $|\mu|$.  For a TI with finite Schr{\"o}dinger contribution, the red curve ($\mu>0$) is always above the relativistic case while the dashed blue curve ($\mu<0$) is always below it.  In frame (b), we present the complimentary results for the $|\mu|$ dependence of the cyclotron resonance frequency $\wcr$ for $B=0.1$T.  Again, we see deviations from the pure relativistic limit with the solid red curve ($\mu>0$) and dash-dotted blue curve ($\mu<0$) falling above and below this case (dash-dotted green), respectively.  It is interesting to obtain a first correction to $W_{CR}$ and $\wcr$ as a function of $1/m$.  Trivially, the $m\rightarrow\infty$ limit of Eqn.~\eqref{wcr} gives the well known result for $\wcr$ in the pure relativistic regime; that is, $\wcr=\hbar eB v_F^2/|\mu|$.  The first correction is a factor of
\begin{align}
\wcr^1(\mu)=1+\frac{3\mu }{2mv_F^2}
\end{align}
which, as seen in the Fig.~\ref{fig:Cyc}(b), moves the curve up for $\mu>0$ and down for $\mu<0$.  In Fig.~\ref{fig:Cyc}(b), only $|\mu|>10$ meV is shown as cyclotron resonance is defined for large $\mu$.  For the optical spectral weight, the first correction gives
\begin{align}
W_{CR}\approx\frac{e^2}{8\hbar}\mu\left(1+\frac{\mu }{2mv_F^2}\right).
\end{align} 
Again, the additional contribution to the relativistic limit is positive (negative) for $\mu>0$ ($\mu<0$).

\begin{figure}[h!]
\begin{center}
\includegraphics[width=1.0\linewidth]{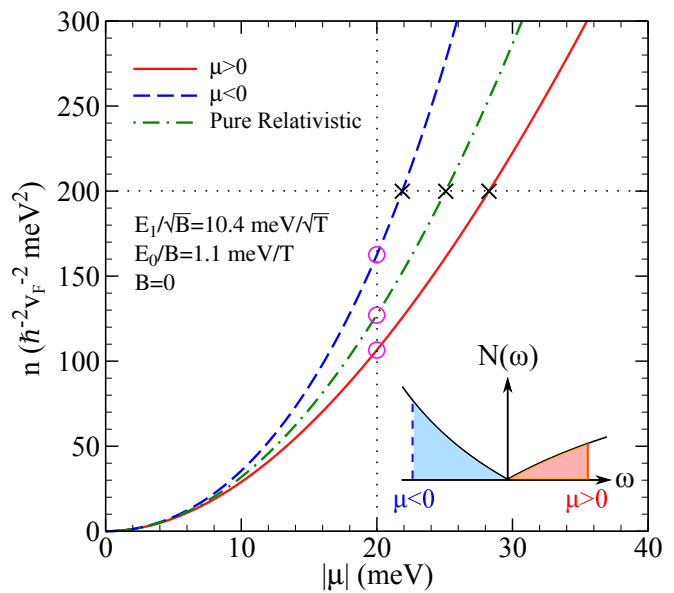}
\end{center}
\caption{\label{fig:density}(Color online) The number of dopants ($n$) as a function of the magnitude of the chemical potential.  The pure relativistic case is contrasted with that when a subdominant Schr\"odinger term is included.  Inset: density of states as a function of $\omega$.  The shading identifies the conduction (red) and valence (blue) bands.
}
\end{figure}

So far, the results have been presented as a function of chemical potential.  Of course, the electronic density of states in a TI is not symmetric with respect to the neutrality (Dirac) point.  It rises above the pure Dirac limit for negative energies and is reduced below this regime for positive $\mu$.  This is shown in the inset of Fig.~\ref{fig:density}.
Thus, for a given magnitude of $\mu$, there are more holes ($n_h$) than electrons ($n_e$) involved.  It is of interest to see how large the asymmetry in $W_{CR}$ and $\wcr$ are when plotted with respect to an equal value of doping (holes or electrons).  To accomplish this conversion, we use the relationship between $\mu$ and $n$.  Namely, 
\begin{align}\label{mu-n}
\sqrt{\frac{4\pi n}{(2mv_F^2)^2}}=\left|1-\sqrt{1+\frac{2\mu}{mv_F^2}}\right|.
\end{align}
Using Eqn.~\eqref{mu-n}, we find
\begin{align}\label{WCR-n}
W_{CR}=e^2\frac{\sqrt{4\pi n}}{16}\left(1\pm\frac{\sqrt{4\pi n}}{2mv_F^2}\right),
\end{align}
and
\begin{align}\label{wcr-n}
\wcr=\frac{eB\hbar v_F^2}{2|\mu|}\left(2\pm\frac{\sqrt{4\pi n}}{2mv_F^2}\right)\left(1\pm\frac{\sqrt{4\pi n}}{2mv_F^2}\right),
\end{align}
for the optical spectral weight and frequency of the cyclotron resonance line, respectively.  In both cases, $\pm$ corresponds to electrons and holes, respectively.  Here,
\begin{align}
|\mu|=\frac{mv_F^2}{2}\left|\left[1\pm \frac{\sqrt{4\pi n}}{2mv_F^2}\right]^2-1\right|.
\end{align}

\begin{figure}[h!]
\begin{center}
\includegraphics[width=1.0\linewidth]{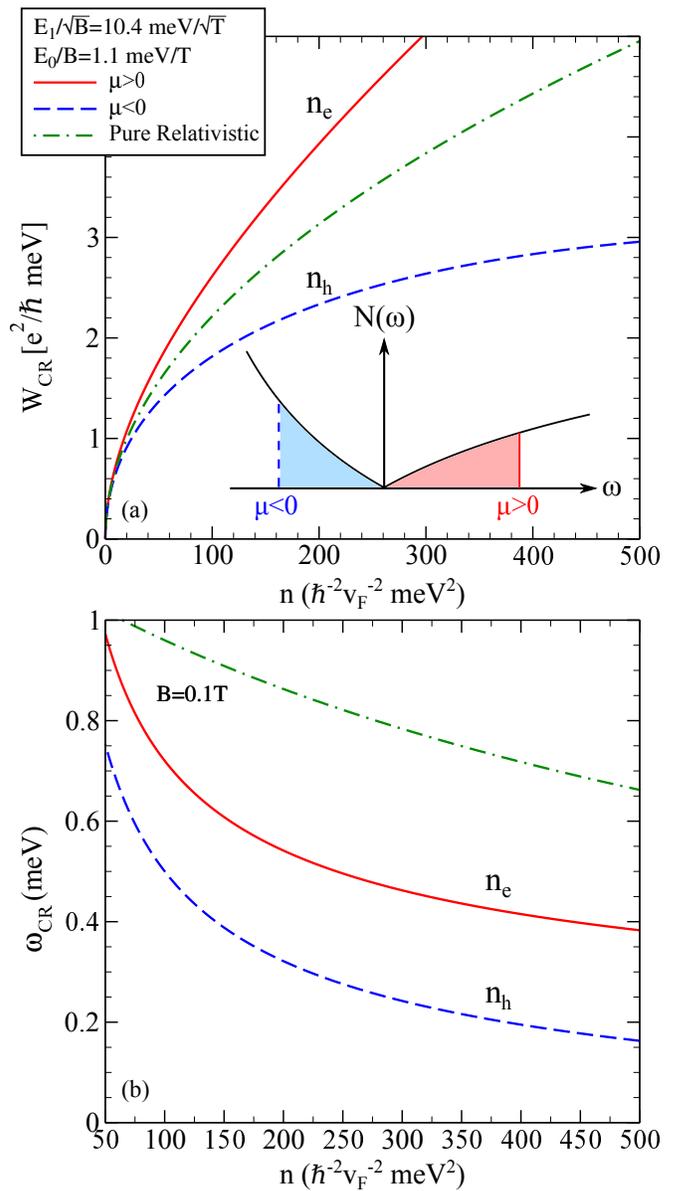}
\end{center}
\caption{\label{fig:Cyc-n}(Color online) (a) Optical spectral weight under the cyclotron resonance line as a function of the number of dopants $n$ for the pure relativistic and TI regimes.  Inset: density of states as a function of $\omega$.  The valence and conduction portions are given by the blue and red shading, respectively. (b) The corresponding results for the cyclotron frequency at $B=0.1$T.
}
\end{figure}

The main frame of Fig.~\ref{fig:density} shows the $\mu$ dependence of the density $n$ in units of $\hbar^{-2}v_F^{-2}$ meV$^2$ for the pure relativistic case (dash-dotted green) and TI with $\mu>0$ (solid red) and $\mu<0$ (dashed blue).  The intersection of the curves with the dotted vertical line at $|\mu|=20$ meV (circles) emphasizes the different electron densities at fixed $\mu$.  The horizontal dotted line gives the different chemical potentials (marked by a $\times$) associated with a fixed particle number ($n=200$ in our units).  As expected, for fixed $\mu$, there are more holes than electrons.  To get a constant number of dopants, the magnitude of the chemical potential needs to be considerably larger in the electron case than for holes.  The impact of this conversion from chemical potential to dopant density is shown in Fig.~\ref{fig:Cyc-n} for the optical spectral weight [frame (a)] and cyclotron resonance frequency [frame (b)].  
Large differences between electrons and holes are noted.

\begin{figure}[h!]
\begin{center}
\includegraphics[width=1.0\linewidth]{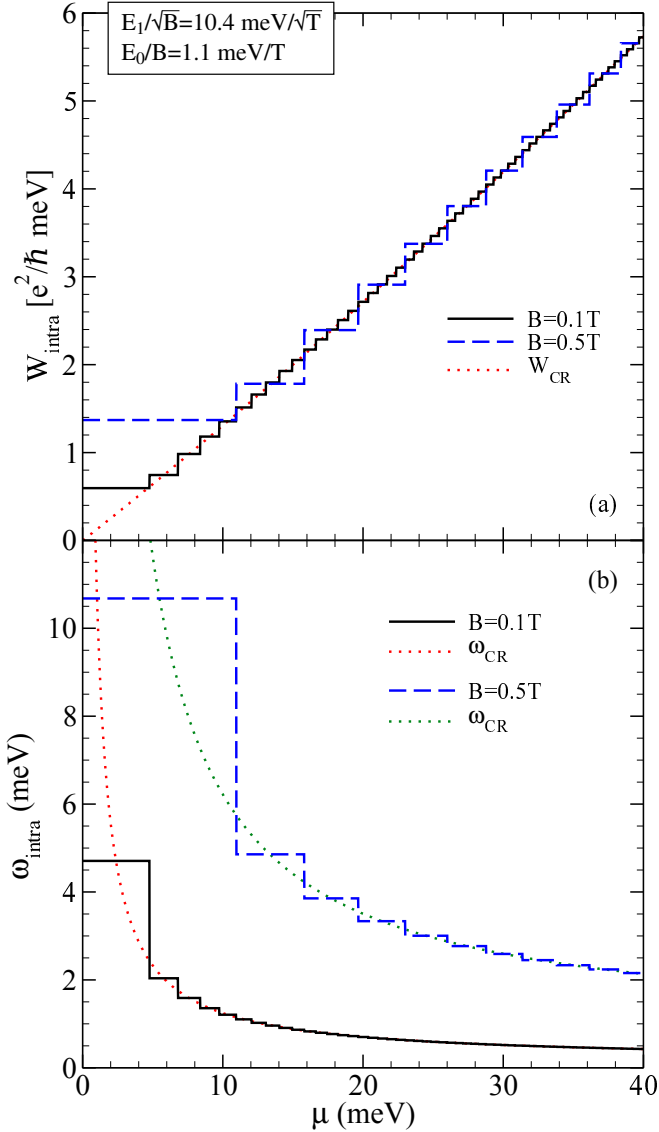}
\end{center}
\caption{\label{fig:Intra}(Color online) (a) Optical spectral weight under the intraband line as a function of $\mu$ for a TI.  Two values of magnetic field are shown: $B=0.1$T and $B=0.5$T.  The dotted curves give the corresponding results for the cyclotron resonance spectral weight. (b)  The position in energy of the intraband line for the same parameters as above. 
}
\end{figure}

Next, we consider how the optical spectral weight under the intraband line ($W_{\rm intra}$) [i.e. the area under the first peak of Fig.~\ref{fig:Condxx-xy}(a)] and its position in energy ($\omega_{\rm intra}$) at finite $B$ approach the cyclotron resonance results as $B$ is decreased.  This is shown in Fig.~\ref{fig:Intra}.
In the top frame, we reproduce as the dotted red curve, the results previously obtained in Fig.~\ref{fig:Cyc} (solid red) for the spectral weight of the cyclotron resonance line $W_{\rm CR}$ in units of $e^2/\hbar$ meV as a function of chemical potential for $\mu>0$.  This curve is to be compared with the related results for the optical spectral weight $W_{\rm intra}$ under the intraband line.  The dashed blue curve is for $B=0.5$T while the solid black curve corresponds to $B=0.1$T.  As the magnetic field is reduced and $\mu$ is increased, our results for $W_{\rm intra}$ approach the dotted red curve for $W_{\rm CR}$ as we expect.  The two curves should agree in the limit of $\mu$ much larger than the magnetic energy.  For $B=0.1$T, this condition is $\mu\gg 3.3$ meV.  In Fig.~\ref{fig:Intra}(b), we show the energy of the intraband transitions $\omega_{\rm intra}$ in meV as a function of $\mu$ for the same two values of magnetic field: $B=0.5$T (dashed blue) and $B=0.1$T (solid black).  The dotted green and red curves are for comparison and give $\wcr$ vs. $\mu$ for $B=0.5$T and $0.1$T, respectively.  As the value of $\mu$ is increased, the condition that $\mu$ be much greater than the magnetic energy scale is better satisfied and $\omega_{\rm intra}$ merges with $\wcr$.  

\section{The Interband Background}

\begin{figure}[h!]
\begin{center}
\includegraphics[width=1.0\linewidth]{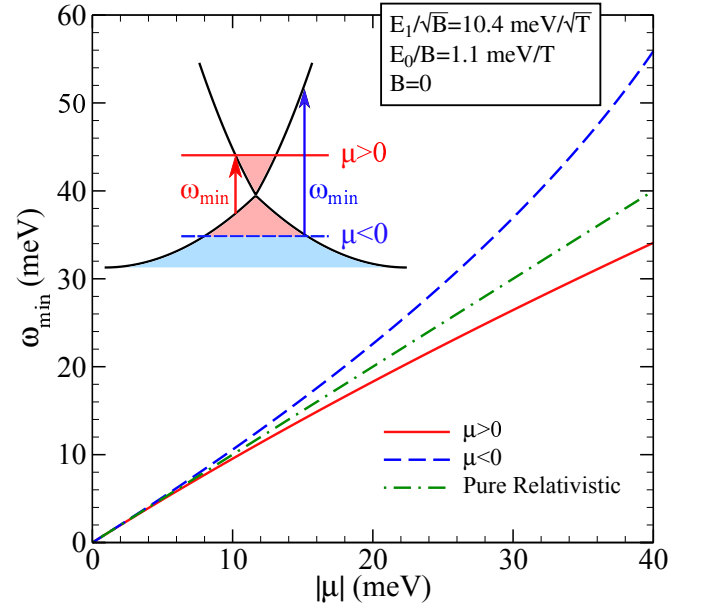}
\end{center}
\caption{\label{fig:wmin}(Color online) Energy cutoff of interband transitions as a function of $|\mu|$.  Positive and negative $\mu$ are considered.  The results are compared to the pure Dirac limit.  Inset: sketch of the dispersion curves.  The arrows identify the interband transitions that define $\omega_{\rm min}$.  Blue corresponds to $\mu<0$ while red is for $\mu>0$.
}
\end{figure}

In the previous section, we emphasized the intraband contribution to the optical conductivity.  However, the interband part is also of interest and corresponds to all but the first magneto-optical lines of Figs.~\ref{fig:Condxx-TI-Dirac}-\ref{fig:Condxx-xy}.  At larger values of photon energy and in the small $B$ limit, these provide the universal background which is given by the known formula
\begin{align}\label{background}
{\rm Re}\sigma_{xx}(\Omega)=\frac{e^2}{16\hbar}\Theta(\Omega-{\rm max}[2\Delta,\omega_{\rm min}]),
\end{align}
where $\omega_{\rm min}$ is the threshold energy for the onset of interband transitions.  In Ref.~\cite{ZLi:2015}, this formula is derived from the form of the conductivity in the case of $B=0$ which involves an integration over momentum.  Eqn.~\eqref{background} properly reduces to the graphene result\cite{Gusynin:2006a,Gusynin:2009} which differs from this equation only in that $\omega_{\rm min}=2\mu$.  In the appendix, we provide an alternative derivation of this formula.  We start with Eqn.~\eqref{sigmaxx} for the longitudinal conductivity at finite $B$ and take the limit of $B\rightarrow 0$.  This serves as a check on our work. The inset of Fig.~\ref{fig:wmin} identifies the energy $\omega_{\rm min}$ for a particular value of $|\mu|$.  
It is smaller for $\mu>0$ (red arrow) than for $\mu<0$ (blue arrow).  The value of $\omega_{\rm min}$ is determined by $\mathcal{E}_+(k_c)-\mathcal{E}_-(k_c)$ with $k_c$ corresponding to $\mu=\mathcal{E}_+(k_c)$ for $\mu>0$ and $\mu=\mathcal{E}_-(k_c)$ for $\mu<0$.   Here, $\mathcal{E}_{\pm}(k)\equiv \hbar^2k^2/(2m)\pm\hbar v_F k$.  This gives
\begin{align}
\omega_{\rm min}(\mu)=2mv_F^2\left|1-\sqrt{1+\frac{2\mu }{mv_F^2}}\right|
\end{align}
which reduces to the known value of $2\mu$ in the pure relativistic limit.  For a TI with a subdominant non-relativistic contribution, our results for $\omega_{\rm min}$ vs. $|\mu|$ are presented in the main frame of Fig.~\ref{fig:wmin}.  The dash-dotted green curve is for comparison ($m\rightarrow\infty$).  For $\mu>0$, the curve (solid red) falls below the green line.  For $\mu<0$, $\omega_{\rm min}$ (dashed blue) is above the relativistic limit.  As we expect, all three curves come together at small $|\mu|$ since, in that limit, the relativistic linear-in-momentum magnetic energy dominates the non-relativistic quadratic term.  As $|\mu|$ is increased, deviations from the pure relativistic result become significant.

\begin{figure}[h!]
\begin{center}
\includegraphics[width=1.0\linewidth]{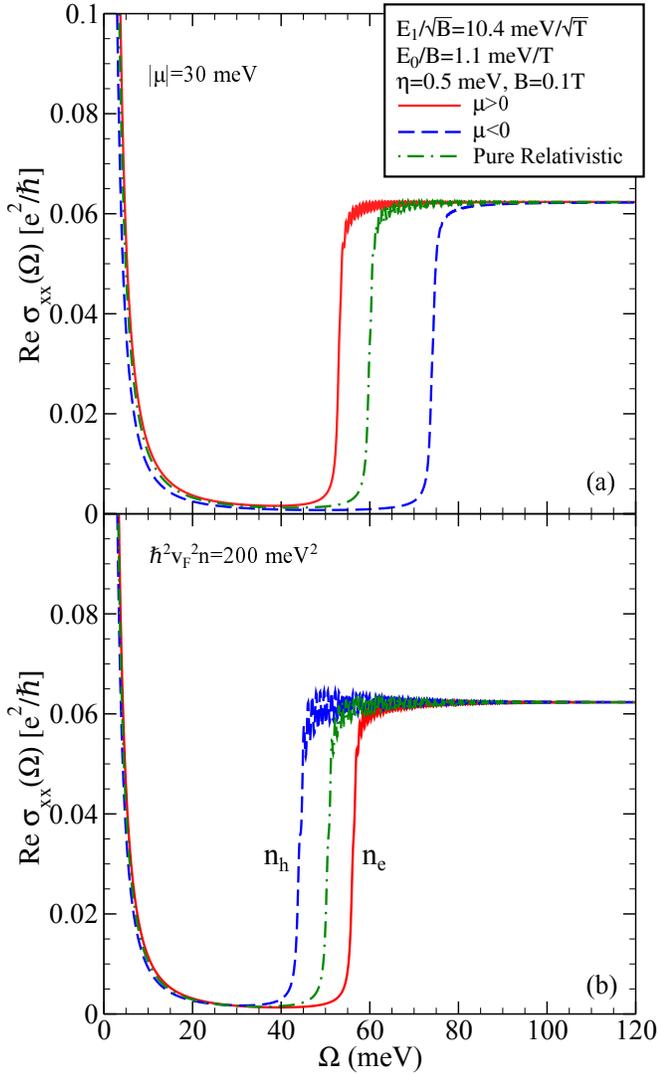}
\end{center}
\caption{\label{fig:Condxx-TI-Dirac-mu}(Color online) Real part of the longitudinal AC conductivity as a function of photon energy.  (a) $|\mu|$ set at 30 meV.  A different interband onset is seen for $\mu\lessgtr 0$ when the Schr\"odinger term is finite. (b) Corresponding results for a fixed dopant concentration $n\hbar^2v_F^2=200$ meV$^2$.
}
\end{figure}

In Fig.~\ref{fig:Condxx-TI-Dirac-mu}(a), we show both the intraband line (Drude-like peak near $\Omega=0$) and the constant interband background which sets in at $\Omega=\omega_{\rm min}$.
The absolute value of $\mu$ is set at 30 meV so that the various onsets for $\mu>0$ (solid red) and $\mu<0$ (dashed blue) in a TI, and the pure Dirac curve (dash-dotted green) are clearly separated.  The height of the background remains at $(1/16)e^2/\hbar$ independent of the mass term.  In frame (b), we show results for fixed dopant density $n$ set at 200$\hbar^{-2} v_F^{-2}$ meV$^2$.  Again, the onset of interband transitions is changed by the subdominant Schr{\"o}dinger term.  Note that these results were obtained from finite magnetic field calculations for $B=0.1$T and consequently small wiggles from the underlying LL structure remain. These wiggles are most pronounced for small values of photon energy.  Since we have included a residual scattering rate $\eta=0.5$ meV, the Drude term centred at $\Omega=0$ is broadened.  It would be a Dirac $\delta$-function in the clean limit.  This smearing parameter is also responsible for the small but finite conductivity seen between the Drude and interband contributions.  Increasing the residual scattering rate would increase the value of this finite conductivity and reduce the small wiggles seen in the interband transitions of Fig.~\ref{fig:Condxx-TI-Dirac-mu}.

\section{Discussion and Conclusions}

We start with the Kubo formula for the real part of the longitudinal and transverse (Hall) conductivities in a magnetic field $B$ which is oriented perpendicular to the surface of a TI.  The magnetic field creates LLs which lead to absorption peaks in the real part of the longitudinal optical conductivity.  The introduction of a subdominant Schr\"odinger contribution to the Hamiltonian is known to split each interband peak into a doublet.  Here, we find that there are associated peaks in the real part of the dynamical Hall conductivity Re$\sigma_{xy}(\Omega)$ which are entirely due to the non-relativistic term.

A primary interest is the semiclassical regime which emerges when the magnitude of the chemical potential $|\mu|$ is much larger than both the relativistic ($E_1$) and non-relativistic ($E_0$) magnetic energy scales.  Nevertheless, we have found it illuminating to begin with a study of the evolution of both the dynamical conductivities Re$\sigma_{xx}(\Omega)$ and Re$\sigma_{xy}(\Omega)$ as the magnitude of the magnetic field is gradually reduced towards the zero-field limit.  For $B=0$, the intraband optical transitions at finite doping provide a Drude contribution to Re$\sigma_{xx}(\Omega)$ and the interband transitions give an additional universal background the height of which is independent of the Schr\"odinger mass; however, its onset is not.  This step is also not particle-hole symmetric whether it be as function of $\mu$ or the number of dopants ($n$).  In the process of considering the interband background, we have provided a new derivation of the relevant formula.  The derivation proceeds from our formula for Re$\sigma_{xx}(\Omega)$ at finite magnetic field and the limit of $B\rightarrow 0$ is formally applied. At finite $B$, we find LL structures associated with the interband transitions in Re$\sigma_{xy}(\Omega)$ which exist only when $m$ is finite.  They are gradually reduced in amplitude with decreasing $B$; for $B=0$, the entire Hall conductivity is zero.  For finite doping, the intraband transitions provide a peak in Re$\sigma_{xx}(\Omega)$ at low photon energy $\Omega$ which evolves into the cyclotron resonance line as $B$ gets small or $|\mu|$ gets large (such that $|\mu|$ is much greater than $E_1$ and $E_0$).  We find a corresponding peak-valley structure in Re$\sigma_{xy}(\Omega)$ with its DC limit equal to the Hall plateau quantization which keeps the pure relativistic limit sequence of $(\pm 1/2, \pm 3/2, \pm 5/2,...)e^2/h$.  The value of $\mu$ at which the transition from one plateau to the next occurs is affected by the non-relativistic mass term.  One consequence of this fact is that at charge neutrality ($\mu=0^+$), the DC Hall conductivity (in units of $e^2/h$) becomes $-1/2$ for $E_0\neq 0$ while it is $1/2$ for $E_0=0$.  This also implies that the structure seen in Re$\sigma_{xy}(\Omega)$ at small $\Omega$ is switched from valley first to peak first.  

When the chemical potential is large compared to the Dirac and Schr{\"o}dinger magnetic scales, the intraband line moves toward the semiclassical cyclotron resonance line.  We provide a simple analytic formula for the cyclotron resonance frequency, its optical spectral weight and its line shape which is found to be unaltered from the pure relativistic case.  However, both the cyclotron frequency ($\wcr$) and optical spectral weight ($W_{CR}$) are changed.  In addition, due to the Schr{\"o}dinger-mass induced particle-hole asymmetry, both $\wcr$ and $W_{CR}$ are different for positive and negative $\mu$.  Differences are also evident for a fixed doping of electrons or holes.  We provide a new derivation of the universal interband background of the longitudinal conductivity.  The work is based on the Kubo formula at finite magnetic field and a formal limit of $B\rightarrow 0$.  We recover the result obtained when no magnetic field is considered.  The height of the background is completely independent of the Schr\"odinger mass $m$ but its onset at $\Omega=\omega_{\rm min}$ is not.  This onset not only depends on $m$ but also depends on the sign of the chemical potential or (for a fixed value of dopant density) whether they are electrons or holes.  

\appendix*

\section{}

In this appendix, we provide a derivation of the universal background formula [Eqn.~\eqref{background}] as the $B\rightarrow 0$ limit of Eqn.~\eqref{sigmaxx}.  The two optical matrix elements of interest are for transitions from ($N-1,-$) to ($N,+$) and ($N,-$) to ($N+1,+$).  In addition, these are only needed for large $N$ with $N=N_c$ given by $\mu=\mathcal{E}_{N_c,+}$.  For simplicity, we have assumed $\mu>0$.  The negative $\mu$ case follows from similar considerations.  We can take $\eta\rightarrow 0$ in Eqn.~\eqref{sigmaxx} to get
\begin{align}\label{sigxx-app}
&{\rm Re}\left\lbrace\frac{\sigma_{xx}(\Omega)}{e^2/\hbar}\right\rbrace=\frac{E_1^2}{2}\sum_{N=0}^\infty\\
&\left\lbrace\frac{f_{N+1,+}-f_{N,-}}{\mathcal{E}_{N,-}-\mathcal{E}_{N+1,+}}\mathcal{F}(N-;N+1+)\delta(\Omega+\mathcal{E}_{N+1,+}-\mathcal{E}_{N,-})\right.\notag\\
&\left.+\frac{f_{N-1,+}-f_{N,-}}{\mathcal{E}_{N,-}-\mathcal{E}_{N-1,+}}\mathcal{F}(N-;N-1+)\delta(\Omega+\mathcal{E}_{N-1,+}-\mathcal{E}_{N,-})\right\rbrace.\notag
\end{align}
In the limit $B\rightarrow 0$ and large $\mu$, the Dirac $\delta$-functions in Eqn.~\eqref{sigxx-app} restrict $N$ to values near $N_c$ where $\Omega=2\sqrt{2N_c}E_1$ which also needs to be larger that $\omega_{\rm min}$.  The $\mathcal{F}$ factors can be replaced by $\mathcal{F}(N_c-;N_c+)$ and the energy denominators by $-2\sqrt{2N_c}E_1$ since $E_0$ goes to zero as $B\rightarrow 0$.  Finally, the thermal factor with a $-$ needs to be replaced by 1 and the one with $+$ by 0.  This allows only transitions with an energy larger than $\omega_{\rm min}$ or $N\geq N_c$.  Thus, we have
\begin{align}
{\rm Re}\left\lbrace\frac{\sigma_{xx}(\Omega)}{e^2/\hbar}\right\rbrace=&\frac{E_1^2}{2}\sum_{N=0}^\infty \frac{1}{\sqrt{2N_c}E_1}\mathcal{F}(N_c-;N_c+)\notag\\
&\times\delta(\Omega+\mathcal{E}_{N,+}-\mathcal{E}_{N,-}),
\end{align}
with $\Omega>\omega_{\rm min}$.  The sum over $N$ can be replaced by an integral over a continuous variable $N$.  Integration is then preformed on the Dirac $\delta$-function which gives a factor of $[(d/dN)2\sqrt{2N}E_1]^{-1}|_{N_c}$.  Therefore,
\begin{align}
{\rm Re}\left\lbrace\frac{\sigma_{xx}(\Omega)}{e^2/\hbar}\right\rbrace=&\frac{E_1^2}{2}\frac{1}{\sqrt{2N_c}E_1}\mathcal{F}(N_c-;N_c+)\frac{\sqrt{2N_c}}{2E_1}.
\end{align}
But, $\mathcal{F}(N_c-;N_c+)=1/4$; so,
\begin{align}
{\rm Re}\sigma_{xx}(\Omega)=\frac{e^2}{16\hbar}
\end{align}
which is completely independent of the Schr\"odinger mass $m$.  However, this constant background starts at $\Omega=\omega_{\rm min}$ and so, as we have seen, this onset does depend on $m$ as emphasized in Eqn.~\eqref{background}.

\begin{acknowledgments}
We thank E. J. Nicol for support of this project.  This work has been supported by the Natural Sciences and Engineering Research Council of Canada and, in part, by the Canadian Institute for Advanced Research.
\end{acknowledgments}

\bibliographystyle{apsrev4-1}
\bibliography{TI-CYC}

\begin{thebibliography}{51}%
\makeatletter
\providecommand \@ifxundefined [1]{%
 \@ifx{#1\undefined}
}%
\providecommand \@ifnum [1]{%
 \ifnum #1\expandafter \@firstoftwo
 \else \expandafter \@secondoftwo
 \fi
}%
\providecommand \@ifx [1]{%
 \ifx #1\expandafter \@firstoftwo
 \else \expandafter \@secondoftwo
 \fi
}%
\providecommand \natexlab [1]{#1}%
\providecommand \enquote  [1]{``#1''}%
\providecommand \bibnamefont  [1]{#1}%
\providecommand \bibfnamefont [1]{#1}%
\providecommand \citenamefont [1]{#1}%
\providecommand \href@noop [0]{\@secondoftwo}%
\providecommand \href [0]{\begingroup \@sanitize@url \@href}%
\providecommand \@href[1]{\@@startlink{#1}\@@href}%
\providecommand \@@href[1]{\endgroup#1\@@endlink}%
\providecommand \@sanitize@url [0]{\catcode `\\12\catcode `\$12\catcode
  `\&12\catcode `\#12\catcode `\^12\catcode `\_12\catcode `\%12\relax}%
\providecommand \@@startlink[1]{}%
\providecommand \@@endlink[0]{}%
\providecommand \url  [0]{\begingroup\@sanitize@url \@url }%
\providecommand \@url [1]{\endgroup\@href {#1}{\urlprefix }}%
\providecommand \urlprefix  [0]{URL }%
\providecommand \Eprint [0]{\href }%
\providecommand \doibase [0]{http://dx.doi.org/}%
\providecommand \selectlanguage [0]{\@gobble}%
\providecommand \bibinfo  [0]{\@secondoftwo}%
\providecommand \bibfield  [0]{\@secondoftwo}%
\providecommand \translation [1]{[#1]}%
\providecommand \BibitemOpen [0]{}%
\providecommand \bibitemStop [0]{}%
\providecommand \bibitemNoStop [0]{.\EOS\space}%
\providecommand \EOS [0]{\spacefactor3000\relax}%
\providecommand \BibitemShut  [1]{\csname bibitem#1\endcsname}%
\let\auto@bib@innerbib\@empty
%</preamble>
\bibitem [{\citenamefont {Kane}\ and\ \citenamefont {Mele}(2005)}]{Kane:2005}%
  \BibitemOpen
  \bibfield  {author} {\bibinfo {author} {\bibfnamefont {C.~L.}\ \bibnamefont
  {Kane}}\ and\ \bibinfo {author} {\bibfnamefont {E.~J.}\ \bibnamefont
  {Mele}},\ }\href@noop {} {\bibfield  {journal} {\bibinfo  {journal} {Phys.
  Rev. Lett.}\ }\textbf {\bibinfo {volume} {95}},\ \bibinfo {pages} {226801}
  (\bibinfo {year} {2005})}\BibitemShut {NoStop}%
\bibitem [{\citenamefont {Hasan}\ and\ \citenamefont
  {Kane}(2010)}]{Hasan:2010}%
  \BibitemOpen
  \bibfield  {author} {\bibinfo {author} {\bibfnamefont {M.~Z.}\ \bibnamefont
  {Hasan}}\ and\ \bibinfo {author} {\bibfnamefont {C.~L.}\ \bibnamefont
  {Kane}},\ }\href {\doibase 10.1103/RevModPhys.82.3045} {\bibfield  {journal}
  {\bibinfo  {journal} {Rev. Mod. Phys.}\ }\textbf {\bibinfo {volume} {82}},\
  \bibinfo {pages} {3045} (\bibinfo {year} {2010})}\BibitemShut {NoStop}%
\bibitem [{\citenamefont {Qi}\ and\ \citenamefont {Zhang}(2011)}]{Qi:2011}%
  \BibitemOpen
  \bibfield  {author} {\bibinfo {author} {\bibfnamefont {X.-L.}\ \bibnamefont
  {Qi}}\ and\ \bibinfo {author} {\bibfnamefont {S.-C.}\ \bibnamefont {Zhang}},\
  }\href {\doibase 10.1103/RevModPhys.83.1057} {\bibfield  {journal} {\bibinfo
  {journal} {Rev. Mod. Phys.}\ }\textbf {\bibinfo {volume} {83}},\ \bibinfo
  {pages} {1057} (\bibinfo {year} {2011})}\BibitemShut {NoStop}%
\bibitem [{\citenamefont {Moore}(2010)}]{Moore:2010}%
  \BibitemOpen
  \bibfield  {author} {\bibinfo {author} {\bibfnamefont {J.~E.}\ \bibnamefont
  {Moore}},\ }\href@noop {} {\bibfield  {journal} {\bibinfo  {journal}
  {Nature}\ }\textbf {\bibinfo {volume} {464}},\ \bibinfo {pages} {194}
  (\bibinfo {year} {2010})}\BibitemShut {NoStop}%
\bibitem [{\citenamefont {Moore}\ and\ \citenamefont
  {Balents}(2007)}]{Moore:2007}%
  \BibitemOpen
  \bibfield  {author} {\bibinfo {author} {\bibfnamefont {J.~E.}\ \bibnamefont
  {Moore}}\ and\ \bibinfo {author} {\bibfnamefont {L.}~\bibnamefont
  {Balents}},\ }\href@noop {} {\bibfield  {journal} {\bibinfo  {journal} {Phys.
  Rev. B}\ }\textbf {\bibinfo {volume} {75}},\ \bibinfo {pages} {121306(R)}
  (\bibinfo {year} {2007})}\BibitemShut {NoStop}%
\bibitem [{\citenamefont {Fu}\ \emph {et~al.}(2007)\citenamefont {Fu},
  \citenamefont {Kane},\ and\ \citenamefont {Mele}}]{Fu:2007}%
  \BibitemOpen
  \bibfield  {author} {\bibinfo {author} {\bibfnamefont {L.}~\bibnamefont
  {Fu}}, \bibinfo {author} {\bibfnamefont {C.~L.}\ \bibnamefont {Kane}}, \ and\
  \bibinfo {author} {\bibfnamefont {E.~J.}\ \bibnamefont {Mele}},\ }\href@noop
  {} {\bibfield  {journal} {\bibinfo  {journal} {Phys. Rev. Lett.}\ }\textbf
  {\bibinfo {volume} {98}},\ \bibinfo {pages} {106803} (\bibinfo {year}
  {2007})}\BibitemShut {NoStop}%
\bibitem [{\citenamefont {Hsieh}\ \emph {et~al.}(2008)\citenamefont {Hsieh},
  \citenamefont {Qian}, \citenamefont {Wray}, \citenamefont {Xia},
  \citenamefont {Hor}, \citenamefont {Cava},\ and\ \citenamefont
  {Hasan}}]{Hsieh:2008}%
  \BibitemOpen
  \bibfield  {author} {\bibinfo {author} {\bibfnamefont {D.}~\bibnamefont
  {Hsieh}}, \bibinfo {author} {\bibfnamefont {D.}~\bibnamefont {Qian}},
  \bibinfo {author} {\bibfnamefont {L.}~\bibnamefont {Wray}}, \bibinfo {author}
  {\bibfnamefont {Y.}~\bibnamefont {Xia}}, \bibinfo {author} {\bibfnamefont
  {Y.~S.}\ \bibnamefont {Hor}}, \bibinfo {author} {\bibfnamefont {R.~J.}\
  \bibnamefont {Cava}}, \ and\ \bibinfo {author} {\bibfnamefont {M.~Z.}\
  \bibnamefont {Hasan}},\ }\href@noop {} {\bibfield  {journal} {\bibinfo
  {journal} {Nature}\ }\textbf {\bibinfo {volume} {452}},\ \bibinfo {pages}
  {970} (\bibinfo {year} {2008})}\BibitemShut {NoStop}%
\bibitem [{\citenamefont {Fu}\ and\ \citenamefont {Kane}(2008)}]{Fu:2008}%
  \BibitemOpen
  \bibfield  {author} {\bibinfo {author} {\bibfnamefont {L.}~\bibnamefont
  {Fu}}\ and\ \bibinfo {author} {\bibfnamefont {C.~L.}\ \bibnamefont {Kane}},\
  }\href {\doibase 10.1103/PhysRevLett.100.096407} {\bibfield  {journal}
  {\bibinfo  {journal} {Phys. Rev. Lett.}\ }\textbf {\bibinfo {volume} {100}},\
  \bibinfo {pages} {096407} (\bibinfo {year} {2008})}\BibitemShut {NoStop}%
\bibitem [{\citenamefont {Zhang}\ \emph {et~al.}(2009)\citenamefont {Zhang},
  \citenamefont {Liu}, \citenamefont {Qi}, \citenamefont {Dai}, \citenamefont
  {Fang},\ and\ \citenamefont {Zhang}}]{Zhang:2009}%
  \BibitemOpen
  \bibfield  {author} {\bibinfo {author} {\bibfnamefont {H.-J.}\ \bibnamefont
  {Zhang}}, \bibinfo {author} {\bibfnamefont {C.-X.}\ \bibnamefont {Liu}},
  \bibinfo {author} {\bibfnamefont {X.-L.}\ \bibnamefont {Qi}}, \bibinfo
  {author} {\bibfnamefont {X.}~\bibnamefont {Dai}}, \bibinfo {author}
  {\bibfnamefont {Z.}~\bibnamefont {Fang}}, \ and\ \bibinfo {author}
  {\bibfnamefont {S.-C.}\ \bibnamefont {Zhang}},\ }\href@noop {} {\bibfield
  {journal} {\bibinfo  {journal} {Nature Phys.}\ }\textbf {\bibinfo {volume}
  {5}},\ \bibinfo {pages} {438} (\bibinfo {year} {2009})}\BibitemShut {NoStop}%
\bibitem [{\citenamefont {Ando}(2013)}]{Ando:2013}%
  \BibitemOpen
  \bibfield  {author} {\bibinfo {author} {\bibfnamefont {Y.}~\bibnamefont
  {Ando}},\ }\href@noop {} {\bibfield  {journal} {\bibinfo  {journal} {J. Phys.
  Soc. Jpn.}\ }\textbf {\bibinfo {volume} {82}},\ \bibinfo {pages} {102001}
  (\bibinfo {year} {2013})}\BibitemShut {NoStop}%
\bibitem [{\citenamefont {Hsieh}\ \emph
  {et~al.}(2009{\natexlab{a}})\citenamefont {Hsieh}, \citenamefont {Xia},
  \citenamefont {Qian}, \citenamefont {Wray}, \citenamefont {Dil},
  \citenamefont {Meier}, \citenamefont {Osterwalder}, \citenamefont {Patthey},
  \citenamefont {Checkelsky}, \citenamefont {Ong}, \citenamefont {Fedorov},
  \citenamefont {Lin}, \citenamefont {Bansil}, \citenamefont {Grauer},
  \citenamefont {Hor}, \citenamefont {Cava},\ and\ \citenamefont
  {Hasan}}]{Hsieh:2009}%
  \BibitemOpen
  \bibfield  {author} {\bibinfo {author} {\bibfnamefont {D.}~\bibnamefont
  {Hsieh}}, \bibinfo {author} {\bibfnamefont {Y.}~\bibnamefont {Xia}}, \bibinfo
  {author} {\bibfnamefont {D.}~\bibnamefont {Qian}}, \bibinfo {author}
  {\bibfnamefont {L.}~\bibnamefont {Wray}}, \bibinfo {author} {\bibfnamefont
  {J.~H.}\ \bibnamefont {Dil}}, \bibinfo {author} {\bibfnamefont
  {F.}~\bibnamefont {Meier}}, \bibinfo {author} {\bibfnamefont
  {J.}~\bibnamefont {Osterwalder}}, \bibinfo {author} {\bibfnamefont
  {L.}~\bibnamefont {Patthey}}, \bibinfo {author} {\bibfnamefont {J.~G.}\
  \bibnamefont {Checkelsky}}, \bibinfo {author} {\bibfnamefont {N.~P.}\
  \bibnamefont {Ong}}, \bibinfo {author} {\bibfnamefont {A.~V.}\ \bibnamefont
  {Fedorov}}, \bibinfo {author} {\bibfnamefont {H.}~\bibnamefont {Lin}},
  \bibinfo {author} {\bibfnamefont {A.}~\bibnamefont {Bansil}}, \bibinfo
  {author} {\bibfnamefont {D.}~\bibnamefont {Grauer}}, \bibinfo {author}
  {\bibfnamefont {Y.~S.}\ \bibnamefont {Hor}}, \bibinfo {author} {\bibfnamefont
  {R.~J.}\ \bibnamefont {Cava}}, \ and\ \bibinfo {author} {\bibfnamefont
  {M.~Z.}\ \bibnamefont {Hasan}},\ }\href@noop {} {\bibfield  {journal}
  {\bibinfo  {journal} {Nature}\ }\textbf {\bibinfo {volume} {460}},\ \bibinfo
  {pages} {1101} (\bibinfo {year} {2009}{\natexlab{a}})}\BibitemShut {NoStop}%
\bibitem [{\citenamefont {Hsieh}\ \emph
  {et~al.}(2009{\natexlab{b}})\citenamefont {Hsieh}, \citenamefont {Xia},
  \citenamefont {Wray}, \citenamefont {Qian}, \citenamefont {Pal},
  \citenamefont {Dil}, \citenamefont {Osterwalder}, \citenamefont {Meier},
  \citenamefont {Bihlmayer}, \citenamefont {Kane}, \citenamefont {Hor},
  \citenamefont {Cava},\ and\ \citenamefont {Hasan}}]{Hsieh:2009a}%
  \BibitemOpen
  \bibfield  {author} {\bibinfo {author} {\bibfnamefont {D.}~\bibnamefont
  {Hsieh}}, \bibinfo {author} {\bibfnamefont {Y.}~\bibnamefont {Xia}}, \bibinfo
  {author} {\bibfnamefont {L.}~\bibnamefont {Wray}}, \bibinfo {author}
  {\bibfnamefont {D.}~\bibnamefont {Qian}}, \bibinfo {author} {\bibfnamefont
  {A.}~\bibnamefont {Pal}}, \bibinfo {author} {\bibfnamefont {J.~H.}\
  \bibnamefont {Dil}}, \bibinfo {author} {\bibfnamefont {J.}~\bibnamefont
  {Osterwalder}}, \bibinfo {author} {\bibfnamefont {F.}~\bibnamefont {Meier}},
  \bibinfo {author} {\bibfnamefont {G.}~\bibnamefont {Bihlmayer}}, \bibinfo
  {author} {\bibfnamefont {C.~L.}\ \bibnamefont {Kane}}, \bibinfo {author}
  {\bibfnamefont {Y.~S.}\ \bibnamefont {Hor}}, \bibinfo {author} {\bibfnamefont
  {R.~J.}\ \bibnamefont {Cava}}, \ and\ \bibinfo {author} {\bibfnamefont
  {M.~Z.}\ \bibnamefont {Hasan}},\ }\href@noop {} {\bibfield  {journal}
  {\bibinfo  {journal} {Science}\ }\textbf {\bibinfo {volume} {323}},\ \bibinfo
  {pages} {919} (\bibinfo {year} {2009}{\natexlab{b}})}\BibitemShut {NoStop}%
\bibitem [{\citenamefont {Geim}\ and\ \citenamefont
  {Novoselov}(2007)}]{Geim:2007}%
  \BibitemOpen
  \bibfield  {author} {\bibinfo {author} {\bibfnamefont {A.~K.}\ \bibnamefont
  {Geim}}\ and\ \bibinfo {author} {\bibfnamefont {K.~S.}\ \bibnamefont
  {Novoselov}},\ }\href@noop {} {\bibfield  {journal} {\bibinfo  {journal}
  {Nature Materials}\ }\textbf {\bibinfo {volume} {6}},\ \bibinfo {pages} {183}
  (\bibinfo {year} {2007})}\BibitemShut {NoStop}%
\bibitem [{\citenamefont {{Castro Neto}}\ \emph {et~al.}(2009)\citenamefont
  {{Castro Neto}}, \citenamefont {Guinea}, \citenamefont {Peres}, \citenamefont
  {Novoselov},\ and\ \citenamefont {Geim}}]{Neto:2009}%
  \BibitemOpen
  \bibfield  {author} {\bibinfo {author} {\bibfnamefont {A.~H.}\ \bibnamefont
  {{Castro Neto}}}, \bibinfo {author} {\bibfnamefont {F.}~\bibnamefont
  {Guinea}}, \bibinfo {author} {\bibfnamefont {N.~M.~R.}\ \bibnamefont
  {Peres}}, \bibinfo {author} {\bibfnamefont {K.~S.}\ \bibnamefont
  {Novoselov}}, \ and\ \bibinfo {author} {\bibfnamefont {A.~K.}\ \bibnamefont
  {Geim}},\ }\href@noop {} {\bibfield  {journal} {\bibinfo  {journal} {Rev.
  Mod. Phys.}\ }\textbf {\bibinfo {volume} {81}},\ \bibinfo {pages} {109}
  (\bibinfo {year} {2009})}\BibitemShut {NoStop}%
\bibitem [{\citenamefont {Li}\ and\ \citenamefont {Carbotte}(2012)}]{ZLi:2012}%
  \BibitemOpen
  \bibfield  {author} {\bibinfo {author} {\bibfnamefont {Z.}~\bibnamefont
  {Li}}\ and\ \bibinfo {author} {\bibfnamefont {J.~P.}\ \bibnamefont
  {Carbotte}},\ }\href@noop {} {\bibfield  {journal} {\bibinfo  {journal}
  {Phys. Rev. B}\ }\textbf {\bibinfo {volume} {86}},\ \bibinfo {pages} {205425}
  (\bibinfo {year} {2012})}\BibitemShut {NoStop}%
\bibitem [{\citenamefont {Stille}\ \emph {et~al.}(2012)\citenamefont {Stille},
  \citenamefont {Tabert},\ and\ \citenamefont {Nicol}}]{Stille:2012}%
  \BibitemOpen
  \bibfield  {author} {\bibinfo {author} {\bibfnamefont {L.}~\bibnamefont
  {Stille}}, \bibinfo {author} {\bibfnamefont {C.~J.}\ \bibnamefont {Tabert}},
  \ and\ \bibinfo {author} {\bibfnamefont {E.~J.}\ \bibnamefont {Nicol}},\
  }\href {\doibase 10.1103/PhysRevB.86.195405} {\bibfield  {journal} {\bibinfo
  {journal} {Phys. Rev. B}\ }\textbf {\bibinfo {volume} {86}},\ \bibinfo
  {pages} {195405} (\bibinfo {year} {2012})}\BibitemShut {NoStop}%
\bibitem [{\citenamefont {Tabert}\ and\ \citenamefont
  {Nicol}(2013)}]{Tabert:2013a}%
  \BibitemOpen
  \bibfield  {author} {\bibinfo {author} {\bibfnamefont {C.~J.}\ \bibnamefont
  {Tabert}}\ and\ \bibinfo {author} {\bibfnamefont {E.~J.}\ \bibnamefont
  {Nicol}},\ }\href {\doibase 10.1103/PhysRevLett.110.197402} {\bibfield
  {journal} {\bibinfo  {journal} {Phys. Rev. Lett.}\ }\textbf {\bibinfo
  {volume} {110}},\ \bibinfo {pages} {197402} (\bibinfo {year}
  {2013})}\BibitemShut {NoStop}%
\bibitem [{\citenamefont {Liu}\ \emph {et~al.}(2010)\citenamefont {Liu},
  \citenamefont {Qi}, \citenamefont {Zhang}, \citenamefont {Dai}, \citenamefont
  {Fang},\ and\ \citenamefont {Zhang}}]{Liu:2010}%
  \BibitemOpen
  \bibfield  {author} {\bibinfo {author} {\bibfnamefont {C.-X.}\ \bibnamefont
  {Liu}}, \bibinfo {author} {\bibfnamefont {X.-L.}\ \bibnamefont {Qi}},
  \bibinfo {author} {\bibfnamefont {H.}~\bibnamefont {Zhang}}, \bibinfo
  {author} {\bibfnamefont {X.}~\bibnamefont {Dai}}, \bibinfo {author}
  {\bibfnamefont {Z.}~\bibnamefont {Fang}}, \ and\ \bibinfo {author}
  {\bibfnamefont {S.-C.}\ \bibnamefont {Zhang}},\ }\href@noop {} {\bibfield
  {journal} {\bibinfo  {journal} {Phys. Rev. B}\ }\textbf {\bibinfo {volume}
  {82}},\ \bibinfo {pages} {045122} (\bibinfo {year} {2010})}\BibitemShut
  {NoStop}%
\bibitem [{\citenamefont {Li}\ and\ \citenamefont {Carbotte}(2013)}]{ZLi:2013}%
  \BibitemOpen
  \bibfield  {author} {\bibinfo {author} {\bibfnamefont {Z.}~\bibnamefont
  {Li}}\ and\ \bibinfo {author} {\bibfnamefont {J.~P.}\ \bibnamefont
  {Carbotte}},\ }\href@noop {} {\bibfield  {journal} {\bibinfo  {journal}
  {Phys. Rev. B}\ }\textbf {\bibinfo {volume} {88}},\ \bibinfo {pages} {045414}
  (\bibinfo {year} {2013})}\BibitemShut {NoStop}%
\bibitem [{\citenamefont {Fuchs}\ \emph {et~al.}(2010)\citenamefont {Fuchs},
  \citenamefont {Piechon}, \citenamefont {Goerbig},\ and\ \citenamefont
  {Montambaux}}]{Fuchs:2010}%
  \BibitemOpen
  \bibfield  {author} {\bibinfo {author} {\bibfnamefont {J.}~\bibnamefont
  {Fuchs}}, \bibinfo {author} {\bibfnamefont {F.}~\bibnamefont {Piechon}},
  \bibinfo {author} {\bibfnamefont {M.}~\bibnamefont {Goerbig}}, \ and\
  \bibinfo {author} {\bibfnamefont {G.}~\bibnamefont {Montambaux}},\
  }\href@noop {} {\bibfield  {journal} {\bibinfo  {journal} {Eur. Phys. J. B}\
  }\textbf {\bibinfo {volume} {77}},\ \bibinfo {pages} {351} (\bibinfo {year}
  {2010})}\BibitemShut {NoStop}%
\bibitem [{\citenamefont {Wright}\ and\ \citenamefont
  {McKenzie}(2013)}]{Wright:2013}%
  \BibitemOpen
  \bibfield  {author} {\bibinfo {author} {\bibfnamefont {A.~R.}\ \bibnamefont
  {Wright}}\ and\ \bibinfo {author} {\bibfnamefont {R.~H.}\ \bibnamefont
  {McKenzie}},\ }\href {\doibase 10.1103/PhysRevB.87.085411} {\bibfield
  {journal} {\bibinfo  {journal} {Phys. Rev. B}\ }\textbf {\bibinfo {volume}
  {87}},\ \bibinfo {pages} {085411} (\bibinfo {year} {2013})}\BibitemShut
  {NoStop}%
\bibitem [{\citenamefont {Goerbig}\ \emph {et~al.}(2014)\citenamefont
  {Goerbig}, \citenamefont {Montambaux},\ and\ \citenamefont
  {Piechon}}]{Goerbig:2014}%
  \BibitemOpen
  \bibfield  {author} {\bibinfo {author} {\bibfnamefont {M.~O.}\ \bibnamefont
  {Goerbig}}, \bibinfo {author} {\bibfnamefont {G.}~\bibnamefont {Montambaux}},
  \ and\ \bibinfo {author} {\bibfnamefont {F.}~\bibnamefont {Piechon}},\
  }\href@noop {} {\bibfield  {journal} {\bibinfo  {journal} {Europhys. Lett.}\
  }\textbf {\bibinfo {volume} {105}},\ \bibinfo {pages} {57005} (\bibinfo
  {year} {2014})}\BibitemShut {NoStop}%
\bibitem [{\citenamefont {Tabert}\ and\ \citenamefont
  {Carbotte}(2015{\natexlab{a}})}]{Tabert:2015}%
  \BibitemOpen
  \bibfield  {author} {\bibinfo {author} {\bibfnamefont {C.~J.}\ \bibnamefont
  {Tabert}}\ and\ \bibinfo {author} {\bibfnamefont {J.~P.}\ \bibnamefont
  {Carbotte}},\ }\href@noop {} {\bibfield  {journal} {\bibinfo  {journal} {J.
  Phys.: Condens. Matter}\ }\textbf {\bibinfo {volume} {27}},\ \bibinfo {pages}
  {015008} (\bibinfo {year} {2015}{\natexlab{a}})}\BibitemShut {NoStop}%
\bibitem [{\citenamefont {Tabert}\ and\ \citenamefont
  {Carbotte}(2015{\natexlab{b}})}]{Tabert:2015b}%
  \BibitemOpen
  \bibfield  {author} {\bibinfo {author} {\bibfnamefont {C.~J.}\ \bibnamefont
  {Tabert}}\ and\ \bibinfo {author} {\bibfnamefont {J.~P.}\ \bibnamefont
  {Carbotte}},\ }\href@noop {} {\bibfield  {journal} {\bibinfo  {journal}
  {Phys. Rev. B}\ }\textbf {\bibinfo {volume} {91}},\ \bibinfo {pages} {235405}
  (\bibinfo {year} {2015}{\natexlab{b}})}\BibitemShut {NoStop}%
\bibitem [{\citenamefont {Taskin}\ and\ \citenamefont
  {Ando}(2011)}]{Taskin:2011a}%
  \BibitemOpen
  \bibfield  {author} {\bibinfo {author} {\bibfnamefont {A.~A.}\ \bibnamefont
  {Taskin}}\ and\ \bibinfo {author} {\bibfnamefont {Y.}~\bibnamefont {Ando}},\
  }\href {\doibase 10.1103/PhysRevB.84.035301} {\bibfield  {journal} {\bibinfo
  {journal} {Phys. Rev. B}\ }\textbf {\bibinfo {volume} {84}},\ \bibinfo
  {pages} {035301} (\bibinfo {year} {2011})}\BibitemShut {NoStop}%
\bibitem [{\citenamefont {LaForge}\ \emph {et~al.}(2010)\citenamefont
  {LaForge}, \citenamefont {Frenzel}, \citenamefont {Pursley}, \citenamefont
  {Lin}, \citenamefont {Liu}, \citenamefont {Shi},\ and\ \citenamefont
  {Basov}}]{LaForge:2010}%
  \BibitemOpen
  \bibfield  {author} {\bibinfo {author} {\bibfnamefont {A.~D.}\ \bibnamefont
  {LaForge}}, \bibinfo {author} {\bibfnamefont {A.}~\bibnamefont {Frenzel}},
  \bibinfo {author} {\bibfnamefont {B.~C.}\ \bibnamefont {Pursley}}, \bibinfo
  {author} {\bibfnamefont {T.}~\bibnamefont {Lin}}, \bibinfo {author}
  {\bibfnamefont {X.}~\bibnamefont {Liu}}, \bibinfo {author} {\bibfnamefont
  {J.}~\bibnamefont {Shi}}, \ and\ \bibinfo {author} {\bibfnamefont {D.~N.}\
  \bibnamefont {Basov}},\ }\href {\doibase 10.1103/PhysRevB.81.125120}
  {\bibfield  {journal} {\bibinfo  {journal} {Phys. Rev. B}\ }\textbf {\bibinfo
  {volume} {81}},\ \bibinfo {pages} {125120} (\bibinfo {year}
  {2010})}\BibitemShut {NoStop}%
\bibitem [{\citenamefont {Xi}\ \emph {et~al.}(2013)\citenamefont {Xi},
  \citenamefont {Ma}, \citenamefont {Liu}, \citenamefont {Chen}, \citenamefont
  {Ku}, \citenamefont {Berger}, \citenamefont {Martin}, \citenamefont
  {Tanner},\ and\ \citenamefont {Carr}}]{Xi:2013}%
  \BibitemOpen
  \bibfield  {author} {\bibinfo {author} {\bibfnamefont {X.}~\bibnamefont
  {Xi}}, \bibinfo {author} {\bibfnamefont {C.}~\bibnamefont {Ma}}, \bibinfo
  {author} {\bibfnamefont {Z.}~\bibnamefont {Liu}}, \bibinfo {author}
  {\bibfnamefont {Z.}~\bibnamefont {Chen}}, \bibinfo {author} {\bibfnamefont
  {W.}~\bibnamefont {Ku}}, \bibinfo {author} {\bibfnamefont {H.}~\bibnamefont
  {Berger}}, \bibinfo {author} {\bibfnamefont {C.}~\bibnamefont {Martin}},
  \bibinfo {author} {\bibfnamefont {D.~B.}\ \bibnamefont {Tanner}}, \ and\
  \bibinfo {author} {\bibfnamefont {G.~L.}\ \bibnamefont {Carr}},\ }\href
  {\doibase 10.1103/PhysRevLett.111.155701} {\bibfield  {journal} {\bibinfo
  {journal} {Phys. Rev. Lett.}\ }\textbf {\bibinfo {volume} {111}},\ \bibinfo
  {pages} {155701} (\bibinfo {year} {2013})}\BibitemShut {NoStop}%
\bibitem [{\citenamefont {Chapler}\ \emph {et~al.}(2014)\citenamefont
  {Chapler}, \citenamefont {Post}, \citenamefont {Richardella}, \citenamefont
  {Lee}, \citenamefont {Tao}, \citenamefont {Samarth},\ and\ \citenamefont
  {Basov}}]{Chapler:2014}%
  \BibitemOpen
  \bibfield  {author} {\bibinfo {author} {\bibfnamefont {B.~C.}\ \bibnamefont
  {Chapler}}, \bibinfo {author} {\bibfnamefont {K.~W.}\ \bibnamefont {Post}},
  \bibinfo {author} {\bibfnamefont {A.~R.}\ \bibnamefont {Richardella}},
  \bibinfo {author} {\bibfnamefont {J.~S.}\ \bibnamefont {Lee}}, \bibinfo
  {author} {\bibfnamefont {J.}~\bibnamefont {Tao}}, \bibinfo {author}
  {\bibfnamefont {N.}~\bibnamefont {Samarth}}, \ and\ \bibinfo {author}
  {\bibfnamefont {D.~N.}\ \bibnamefont {Basov}},\ }\href {\doibase
  10.1103/PhysRevB.89.235308} {\bibfield  {journal} {\bibinfo  {journal} {Phys.
  Rev. B}\ }\textbf {\bibinfo {volume} {89}},\ \bibinfo {pages} {235308}
  (\bibinfo {year} {2014})}\BibitemShut {NoStop}%
\bibitem [{\citenamefont {Sushkov}\ \emph {et~al.}(2010)\citenamefont
  {Sushkov}, \citenamefont {Jenkins}, \citenamefont {Schmadel}, \citenamefont
  {Butch}, \citenamefont {Paglione},\ and\ \citenamefont
  {Drew}}]{Sushkov:2010}%
  \BibitemOpen
  \bibfield  {author} {\bibinfo {author} {\bibfnamefont {A.~B.}\ \bibnamefont
  {Sushkov}}, \bibinfo {author} {\bibfnamefont {G.~S.}\ \bibnamefont
  {Jenkins}}, \bibinfo {author} {\bibfnamefont {D.~C.}\ \bibnamefont
  {Schmadel}}, \bibinfo {author} {\bibfnamefont {N.~P.}\ \bibnamefont {Butch}},
  \bibinfo {author} {\bibfnamefont {J.}~\bibnamefont {Paglione}}, \ and\
  \bibinfo {author} {\bibfnamefont {H.~D.}\ \bibnamefont {Drew}},\ }\href
  {\doibase 10.1103/PhysRevB.82.125110} {\bibfield  {journal} {\bibinfo
  {journal} {Phys. Rev. B}\ }\textbf {\bibinfo {volume} {82}},\ \bibinfo
  {pages} {125110} (\bibinfo {year} {2010})}\BibitemShut {NoStop}%
\bibitem [{\citenamefont {Jenkins}\ \emph {et~al.}(2013)\citenamefont
  {Jenkins}, \citenamefont {Schmadel}, \citenamefont {Sushkov}, \citenamefont
  {Drew}, \citenamefont {Bichler}, \citenamefont {Koblmueller}, \citenamefont
  {Brahlek}, \citenamefont {Bansal},\ and\ \citenamefont {Oh}}]{Jenkins:2013}%
  \BibitemOpen
  \bibfield  {author} {\bibinfo {author} {\bibfnamefont {G.~S.}\ \bibnamefont
  {Jenkins}}, \bibinfo {author} {\bibfnamefont {D.~C.}\ \bibnamefont
  {Schmadel}}, \bibinfo {author} {\bibfnamefont {A.~B.}\ \bibnamefont
  {Sushkov}}, \bibinfo {author} {\bibfnamefont {H.~D.}\ \bibnamefont {Drew}},
  \bibinfo {author} {\bibfnamefont {M.}~\bibnamefont {Bichler}}, \bibinfo
  {author} {\bibfnamefont {G.}~\bibnamefont {Koblmueller}}, \bibinfo {author}
  {\bibfnamefont {M.}~\bibnamefont {Brahlek}}, \bibinfo {author} {\bibfnamefont
  {N.}~\bibnamefont {Bansal}}, \ and\ \bibinfo {author} {\bibfnamefont
  {S.}~\bibnamefont {Oh}},\ }\href {\doibase 10.1103/PhysRevB.87.155126}
  {\bibfield  {journal} {\bibinfo  {journal} {Phys. Rev. B}\ }\textbf {\bibinfo
  {volume} {87}},\ \bibinfo {pages} {155126} (\bibinfo {year}
  {2013})}\BibitemShut {NoStop}%
\bibitem [{\citenamefont {Wu}\ \emph {et~al.}(2015)\citenamefont {Wu},
  \citenamefont {Tse}, \citenamefont {Brahlek}, \citenamefont {Morris},
  \citenamefont {{Vald\'es Aguilar}}, \citenamefont {Koirala}, \citenamefont
  {Oh},\ and\ \citenamefont {Armitage}}]{Wu:2015}%
  \BibitemOpen
  \bibfield  {author} {\bibinfo {author} {\bibfnamefont {L.}~\bibnamefont
  {Wu}}, \bibinfo {author} {\bibfnamefont {W.-K.}\ \bibnamefont {Tse}},
  \bibinfo {author} {\bibfnamefont {M.}~\bibnamefont {Brahlek}}, \bibinfo
  {author} {\bibfnamefont {C.~M.}\ \bibnamefont {Morris}}, \bibinfo {author}
  {\bibfnamefont {R.}~\bibnamefont {{Vald\'es Aguilar}}}, \bibinfo {author}
  {\bibfnamefont {N.}~\bibnamefont {Koirala}}, \bibinfo {author} {\bibfnamefont
  {S.}~\bibnamefont {Oh}}, \ and\ \bibinfo {author} {\bibfnamefont {N.~P.}\
  \bibnamefont {Armitage}},\ }\href@noop {} {} (\bibinfo {year} {2015}),\
  \Eprint {http://arxiv.org/abs/arXiv:1502.04577} {arXiv:1502.04577}
  \BibitemShut {NoStop}%
\bibitem [{\citenamefont {Hancock}\ \emph {et~al.}(2011)\citenamefont
  {Hancock}, \citenamefont {van Mechelen}, \citenamefont {Kuzmenko},
  \citenamefont {van~der Marel}, \citenamefont {Br\"une}, \citenamefont
  {Novik}, \citenamefont {Astakhov}, \citenamefont {Buhmann},\ and\
  \citenamefont {Molenkamp}}]{Hancock:2011}%
  \BibitemOpen
  \bibfield  {author} {\bibinfo {author} {\bibfnamefont {J.~N.}\ \bibnamefont
  {Hancock}}, \bibinfo {author} {\bibfnamefont {J.~L.~M.}\ \bibnamefont {van
  Mechelen}}, \bibinfo {author} {\bibfnamefont {A.~B.}\ \bibnamefont
  {Kuzmenko}}, \bibinfo {author} {\bibfnamefont {D.}~\bibnamefont {van~der
  Marel}}, \bibinfo {author} {\bibfnamefont {C.}~\bibnamefont {Br\"une}},
  \bibinfo {author} {\bibfnamefont {E.~G.}\ \bibnamefont {Novik}}, \bibinfo
  {author} {\bibfnamefont {G.~V.}\ \bibnamefont {Astakhov}}, \bibinfo {author}
  {\bibfnamefont {H.}~\bibnamefont {Buhmann}}, \ and\ \bibinfo {author}
  {\bibfnamefont {L.~W.}\ \bibnamefont {Molenkamp}},\ }\href {\doibase
  10.1103/PhysRevLett.107.136803} {\bibfield  {journal} {\bibinfo  {journal}
  {Phys. Rev. Lett.}\ }\textbf {\bibinfo {volume} {107}},\ \bibinfo {pages}
  {136803} (\bibinfo {year} {2011})}\BibitemShut {NoStop}%
\bibitem [{\citenamefont {Br\"une}\ \emph {et~al.}(2011)\citenamefont
  {Br\"une}, \citenamefont {Liu}, \citenamefont {Novik}, \citenamefont
  {Hankiewicz}, \citenamefont {Buhmann}, \citenamefont {Chen}, \citenamefont
  {Qi}, \citenamefont {Shen}, \citenamefont {Zhang},\ and\ \citenamefont
  {Molenkamp}}]{Brune:2011}%
  \BibitemOpen
  \bibfield  {author} {\bibinfo {author} {\bibfnamefont {C.}~\bibnamefont
  {Br\"une}}, \bibinfo {author} {\bibfnamefont {C.~X.}\ \bibnamefont {Liu}},
  \bibinfo {author} {\bibfnamefont {E.~G.}\ \bibnamefont {Novik}}, \bibinfo
  {author} {\bibfnamefont {E.~M.}\ \bibnamefont {Hankiewicz}}, \bibinfo
  {author} {\bibfnamefont {H.}~\bibnamefont {Buhmann}}, \bibinfo {author}
  {\bibfnamefont {Y.~L.}\ \bibnamefont {Chen}}, \bibinfo {author}
  {\bibfnamefont {X.~L.}\ \bibnamefont {Qi}}, \bibinfo {author} {\bibfnamefont
  {Z.~X.}\ \bibnamefont {Shen}}, \bibinfo {author} {\bibfnamefont {S.~C.}\
  \bibnamefont {Zhang}}, \ and\ \bibinfo {author} {\bibfnamefont {L.~W.}\
  \bibnamefont {Molenkamp}},\ }\href {\doibase 10.1103/PhysRevLett.106.126803}
  {\bibfield  {journal} {\bibinfo  {journal} {Phys. Rev. Lett.}\ }\textbf
  {\bibinfo {volume} {106}},\ \bibinfo {pages} {126803} (\bibinfo {year}
  {2011})}\BibitemShut {NoStop}%
\bibitem [{\citenamefont {Schafgans}\ \emph {et~al.}(2012)\citenamefont
  {Schafgans}, \citenamefont {Post}, \citenamefont {Taskin}, \citenamefont
  {Ando}, \citenamefont {Qi}, \citenamefont {Chapler},\ and\ \citenamefont
  {Basov}}]{Schafgans:2012}%
  \BibitemOpen
  \bibfield  {author} {\bibinfo {author} {\bibfnamefont {A.~A.}\ \bibnamefont
  {Schafgans}}, \bibinfo {author} {\bibfnamefont {K.~W.}\ \bibnamefont {Post}},
  \bibinfo {author} {\bibfnamefont {A.~A.}\ \bibnamefont {Taskin}}, \bibinfo
  {author} {\bibfnamefont {Y.}~\bibnamefont {Ando}}, \bibinfo {author}
  {\bibfnamefont {X.-L.}\ \bibnamefont {Qi}}, \bibinfo {author} {\bibfnamefont
  {B.~C.}\ \bibnamefont {Chapler}}, \ and\ \bibinfo {author} {\bibfnamefont
  {D.~N.}\ \bibnamefont {Basov}},\ }\href {\doibase 10.1103/PhysRevB.85.195440}
  {\bibfield  {journal} {\bibinfo  {journal} {Phys. Rev. B}\ }\textbf {\bibinfo
  {volume} {85}},\ \bibinfo {pages} {195440} (\bibinfo {year}
  {2012})}\BibitemShut {NoStop}%
\bibitem [{\citenamefont {Taskin}\ and\ \citenamefont
  {Ando}(2009)}]{Taskin:2009}%
  \BibitemOpen
  \bibfield  {author} {\bibinfo {author} {\bibfnamefont {A.~A.}\ \bibnamefont
  {Taskin}}\ and\ \bibinfo {author} {\bibfnamefont {Y.}~\bibnamefont {Ando}},\
  }\href@noop {} {\bibfield  {journal} {\bibinfo  {journal} {Phys. Rev. B}\
  }\textbf {\bibinfo {volume} {80}},\ \bibinfo {pages} {085303} (\bibinfo
  {year} {2009})}\BibitemShut {NoStop}%
\bibitem [{\citenamefont {Vald\'es~Aguilar}\ \emph {et~al.}(2015)\citenamefont
  {Vald\'es~Aguilar}, \citenamefont {Qi}, \citenamefont {Brahlek},
  \citenamefont {Bansal}, \citenamefont {Azad}, \citenamefont {Bowlan},
  \citenamefont {Oh}, \citenamefont {Taylor}, \citenamefont {Prasankumar},\
  and\ \citenamefont {Yarotski}}]{Valdes:2015}%
  \BibitemOpen
  \bibfield  {author} {\bibinfo {author} {\bibfnamefont {R.}~\bibnamefont
  {Vald\'es~Aguilar}}, \bibinfo {author} {\bibfnamefont {J.}~\bibnamefont
  {Qi}}, \bibinfo {author} {\bibfnamefont {M.}~\bibnamefont {Brahlek}},
  \bibinfo {author} {\bibfnamefont {N.}~\bibnamefont {Bansal}}, \bibinfo
  {author} {\bibfnamefont {A.}~\bibnamefont {Azad}}, \bibinfo {author}
  {\bibfnamefont {J.}~\bibnamefont {Bowlan}}, \bibinfo {author} {\bibfnamefont
  {S.}~\bibnamefont {Oh}}, \bibinfo {author} {\bibfnamefont {A.~J.}\
  \bibnamefont {Taylor}}, \bibinfo {author} {\bibfnamefont {R.~P.}\
  \bibnamefont {Prasankumar}}, \ and\ \bibinfo {author} {\bibfnamefont {D.~A.}\
  \bibnamefont {Yarotski}},\ }\href {\doibase
  http://dx.doi.org/10.1063/1.4905438} {\bibfield  {journal} {\bibinfo
  {journal} {Applied Physics Letters}\ }\textbf {\bibinfo {volume} {106}},\
  \bibinfo {pages} {011901} (\bibinfo {year} {2015})}\BibitemShut {NoStop}%
\bibitem [{\citenamefont {Crepaldi}\ \emph {et~al.}(2013)\citenamefont
  {Crepaldi}, \citenamefont {Cilento}, \citenamefont {Ressel}, \citenamefont
  {Cacho}, \citenamefont {Johannsen}, \citenamefont {Zacchigna}, \citenamefont
  {Berger}, \citenamefont {Bugnon}, \citenamefont {Grazioli}, \citenamefont
  {Turcu}, \citenamefont {Springate}, \citenamefont {Kern}, \citenamefont
  {Grioni},\ and\ \citenamefont {Parmigiani}}]{Crepaldi:2013}%
  \BibitemOpen
  \bibfield  {author} {\bibinfo {author} {\bibfnamefont {A.}~\bibnamefont
  {Crepaldi}}, \bibinfo {author} {\bibfnamefont {F.}~\bibnamefont {Cilento}},
  \bibinfo {author} {\bibfnamefont {B.}~\bibnamefont {Ressel}}, \bibinfo
  {author} {\bibfnamefont {C.}~\bibnamefont {Cacho}}, \bibinfo {author}
  {\bibfnamefont {J.~C.}\ \bibnamefont {Johannsen}}, \bibinfo {author}
  {\bibfnamefont {M.}~\bibnamefont {Zacchigna}}, \bibinfo {author}
  {\bibfnamefont {H.}~\bibnamefont {Berger}}, \bibinfo {author} {\bibfnamefont
  {P.}~\bibnamefont {Bugnon}}, \bibinfo {author} {\bibfnamefont
  {C.}~\bibnamefont {Grazioli}}, \bibinfo {author} {\bibfnamefont {I.~C.~E.}\
  \bibnamefont {Turcu}}, \bibinfo {author} {\bibfnamefont {E.}~\bibnamefont
  {Springate}}, \bibinfo {author} {\bibfnamefont {K.}~\bibnamefont {Kern}},
  \bibinfo {author} {\bibfnamefont {M.}~\bibnamefont {Grioni}}, \ and\ \bibinfo
  {author} {\bibfnamefont {F.}~\bibnamefont {Parmigiani}},\ }\href {\doibase
  10.1103/PhysRevB.88.121404} {\bibfield  {journal} {\bibinfo  {journal} {Phys.
  Rev. B}\ }\textbf {\bibinfo {volume} {88}},\ \bibinfo {pages} {121404}
  (\bibinfo {year} {2013})}\BibitemShut {NoStop}%
\bibitem [{\citenamefont {Sobota}\ \emph {et~al.}(2014)\citenamefont {Sobota},
  \citenamefont {Yang}, \citenamefont {Leuenberger}, \citenamefont {Kemper},
  \citenamefont {Analytis}, \citenamefont {Fisher}, \citenamefont {Kirchmann},
  \citenamefont {Devereaux},\ and\ \citenamefont {Shen}}]{Sobota:2014}%
  \BibitemOpen
  \bibfield  {author} {\bibinfo {author} {\bibfnamefont {J.~A.}\ \bibnamefont
  {Sobota}}, \bibinfo {author} {\bibfnamefont {S.-L.}\ \bibnamefont {Yang}},
  \bibinfo {author} {\bibfnamefont {D.}~\bibnamefont {Leuenberger}}, \bibinfo
  {author} {\bibfnamefont {A.~F.}\ \bibnamefont {Kemper}}, \bibinfo {author}
  {\bibfnamefont {J.~G.}\ \bibnamefont {Analytis}}, \bibinfo {author}
  {\bibfnamefont {I.~R.}\ \bibnamefont {Fisher}}, \bibinfo {author}
  {\bibfnamefont {P.~S.}\ \bibnamefont {Kirchmann}}, \bibinfo {author}
  {\bibfnamefont {T.~P.}\ \bibnamefont {Devereaux}}, \ and\ \bibinfo {author}
  {\bibfnamefont {Z.-X.}\ \bibnamefont {Shen}},\ }\href {\doibase
  10.1103/PhysRevLett.113.157401} {\bibfield  {journal} {\bibinfo  {journal}
  {Phys. Rev. Lett.}\ }\textbf {\bibinfo {volume} {113}},\ \bibinfo {pages}
  {157401} (\bibinfo {year} {2014})}\BibitemShut {NoStop}%
\bibitem [{\citenamefont {Hatch}\ \emph {et~al.}(2011)\citenamefont {Hatch},
  \citenamefont {Bianchi}, \citenamefont {Guan}, \citenamefont {Bao},
  \citenamefont {Mi}, \citenamefont {Iversen}, \citenamefont {Nilsson},
  \citenamefont {Hornek\ae{}r},\ and\ \citenamefont {Hofmann}}]{Hatch:2011}%
  \BibitemOpen
  \bibfield  {author} {\bibinfo {author} {\bibfnamefont {R.~C.}\ \bibnamefont
  {Hatch}}, \bibinfo {author} {\bibfnamefont {M.}~\bibnamefont {Bianchi}},
  \bibinfo {author} {\bibfnamefont {D.}~\bibnamefont {Guan}}, \bibinfo {author}
  {\bibfnamefont {S.}~\bibnamefont {Bao}}, \bibinfo {author} {\bibfnamefont
  {J.}~\bibnamefont {Mi}}, \bibinfo {author} {\bibfnamefont {B.~B.}\
  \bibnamefont {Iversen}}, \bibinfo {author} {\bibfnamefont {L.}~\bibnamefont
  {Nilsson}}, \bibinfo {author} {\bibfnamefont {L.}~\bibnamefont
  {Hornek\ae{}r}}, \ and\ \bibinfo {author} {\bibfnamefont {P.}~\bibnamefont
  {Hofmann}},\ }\href {\doibase 10.1103/PhysRevB.83.241303} {\bibfield
  {journal} {\bibinfo  {journal} {Phys. Rev. B}\ }\textbf {\bibinfo {volume}
  {83}},\ \bibinfo {pages} {241303} (\bibinfo {year} {2011})}\BibitemShut
  {NoStop}%
\bibitem [{\citenamefont {Pan}\ \emph {et~al.}(2012)\citenamefont {Pan},
  \citenamefont {Fedorov}, \citenamefont {Gardner}, \citenamefont {Lee},
  \citenamefont {Chu},\ and\ \citenamefont {Valla}}]{Pan:2012}%
  \BibitemOpen
  \bibfield  {author} {\bibinfo {author} {\bibfnamefont {Z.-H.}\ \bibnamefont
  {Pan}}, \bibinfo {author} {\bibfnamefont {A.~V.}\ \bibnamefont {Fedorov}},
  \bibinfo {author} {\bibfnamefont {D.}~\bibnamefont {Gardner}}, \bibinfo
  {author} {\bibfnamefont {Y.~S.}\ \bibnamefont {Lee}}, \bibinfo {author}
  {\bibfnamefont {S.}~\bibnamefont {Chu}}, \ and\ \bibinfo {author}
  {\bibfnamefont {T.}~\bibnamefont {Valla}},\ }\href {\doibase
  10.1103/PhysRevLett.108.187001} {\bibfield  {journal} {\bibinfo  {journal}
  {Phys. Rev. Lett.}\ }\textbf {\bibinfo {volume} {108}},\ \bibinfo {pages}
  {187001} (\bibinfo {year} {2012})}\BibitemShut {NoStop}%
\bibitem [{\citenamefont {Kondo}\ \emph {et~al.}(2013)\citenamefont {Kondo},
  \citenamefont {Nakashima}, \citenamefont {Ota}, \citenamefont {Ishida},
  \citenamefont {Malaeb}, \citenamefont {Okazaki}, \citenamefont {Shin},
  \citenamefont {Kriener}, \citenamefont {Sasaki}, \citenamefont {Segawa},\
  and\ \citenamefont {Ando}}]{Kondo:2013}%
  \BibitemOpen
  \bibfield  {author} {\bibinfo {author} {\bibfnamefont {T.}~\bibnamefont
  {Kondo}}, \bibinfo {author} {\bibfnamefont {Y.}~\bibnamefont {Nakashima}},
  \bibinfo {author} {\bibfnamefont {Y.}~\bibnamefont {Ota}}, \bibinfo {author}
  {\bibfnamefont {Y.}~\bibnamefont {Ishida}}, \bibinfo {author} {\bibfnamefont
  {W.}~\bibnamefont {Malaeb}}, \bibinfo {author} {\bibfnamefont
  {K.}~\bibnamefont {Okazaki}}, \bibinfo {author} {\bibfnamefont
  {S.}~\bibnamefont {Shin}}, \bibinfo {author} {\bibfnamefont {M.}~\bibnamefont
  {Kriener}}, \bibinfo {author} {\bibfnamefont {S.}~\bibnamefont {Sasaki}},
  \bibinfo {author} {\bibfnamefont {K.}~\bibnamefont {Segawa}}, \ and\ \bibinfo
  {author} {\bibfnamefont {Y.}~\bibnamefont {Ando}},\ }\href {\doibase
  10.1103/PhysRevLett.110.217601} {\bibfield  {journal} {\bibinfo  {journal}
  {Phys. Rev. Lett.}\ }\textbf {\bibinfo {volume} {110}},\ \bibinfo {pages}
  {217601} (\bibinfo {year} {2013})}\BibitemShut {NoStop}%
\bibitem [{\citenamefont {Brahlek}\ \emph {et~al.}(2014)\citenamefont
  {Brahlek}, \citenamefont {Koirala}, \citenamefont {Salehi}, \citenamefont
  {Bansal},\ and\ \citenamefont {Oh}}]{Brahlek:2014}%
  \BibitemOpen
  \bibfield  {author} {\bibinfo {author} {\bibfnamefont {M.}~\bibnamefont
  {Brahlek}}, \bibinfo {author} {\bibfnamefont {N.}~\bibnamefont {Koirala}},
  \bibinfo {author} {\bibfnamefont {M.}~\bibnamefont {Salehi}}, \bibinfo
  {author} {\bibfnamefont {N.}~\bibnamefont {Bansal}}, \ and\ \bibinfo {author}
  {\bibfnamefont {S.}~\bibnamefont {Oh}},\ }\href {\doibase
  10.1103/PhysRevLett.113.026801} {\bibfield  {journal} {\bibinfo  {journal}
  {Phys. Rev. Lett.}\ }\textbf {\bibinfo {volume} {113}},\ \bibinfo {pages}
  {026801} (\bibinfo {year} {2014})}\BibitemShut {NoStop}%
\bibitem [{\citenamefont {Lawson}\ \emph {et~al.}(2014)\citenamefont {Lawson},
  \citenamefont {Li}, \citenamefont {Yu}, \citenamefont {Asaba}, \citenamefont
  {Tinsman}, \citenamefont {Gao}, \citenamefont {Wang}, \citenamefont {Hor},\
  and\ \citenamefont {Li}}]{Lawson:2014}%
  \BibitemOpen
  \bibfield  {author} {\bibinfo {author} {\bibfnamefont {B.~J.}\ \bibnamefont
  {Lawson}}, \bibinfo {author} {\bibfnamefont {G.}~\bibnamefont {Li}}, \bibinfo
  {author} {\bibfnamefont {F.}~\bibnamefont {Yu}}, \bibinfo {author}
  {\bibfnamefont {T.}~\bibnamefont {Asaba}}, \bibinfo {author} {\bibfnamefont
  {C.}~\bibnamefont {Tinsman}}, \bibinfo {author} {\bibfnamefont
  {T.}~\bibnamefont {Gao}}, \bibinfo {author} {\bibfnamefont {W.}~\bibnamefont
  {Wang}}, \bibinfo {author} {\bibfnamefont {Y.~S.}\ \bibnamefont {Hor}}, \
  and\ \bibinfo {author} {\bibfnamefont {L.}~\bibnamefont {Li}},\ }\href
  {\doibase 10.1103/PhysRevB.90.195141} {\bibfield  {journal} {\bibinfo
  {journal} {Phys. Rev. B}\ }\textbf {\bibinfo {volume} {90}},\ \bibinfo
  {pages} {195141} (\bibinfo {year} {2014})}\BibitemShut {NoStop}%
\bibitem [{\citenamefont {Eguchi}\ \emph {et~al.}(2014)\citenamefont {Eguchi},
  \citenamefont {Kuroda}, \citenamefont {Shirai}, \citenamefont {Kimura},\ and\
  \citenamefont {Shiraishi}}]{Eguchi:2014}%
  \BibitemOpen
  \bibfield  {author} {\bibinfo {author} {\bibfnamefont {G.}~\bibnamefont
  {Eguchi}}, \bibinfo {author} {\bibfnamefont {K.}~\bibnamefont {Kuroda}},
  \bibinfo {author} {\bibfnamefont {K.}~\bibnamefont {Shirai}}, \bibinfo
  {author} {\bibfnamefont {A.}~\bibnamefont {Kimura}}, \ and\ \bibinfo {author}
  {\bibfnamefont {M.}~\bibnamefont {Shiraishi}},\ }\href {\doibase
  10.1103/PhysRevB.90.201307} {\bibfield  {journal} {\bibinfo  {journal} {Phys.
  Rev. B}\ }\textbf {\bibinfo {volume} {90}},\ \bibinfo {pages} {201307}
  (\bibinfo {year} {2014})}\BibitemShut {NoStop}%
\bibitem [{\citenamefont {Kuroda}\ \emph {et~al.}(2010)\citenamefont {Kuroda},
  \citenamefont {Ye}, \citenamefont {Kimura}, \citenamefont {Eremeev},
  \citenamefont {Krasovskii}, \citenamefont {Chulkov}, \citenamefont {Ueda},
  \citenamefont {Miyamoto}, \citenamefont {Okuda}, \citenamefont {Shimada},
  \citenamefont {Namatame},\ and\ \citenamefont {Taniguchi}}]{Kuroda:2010}%
  \BibitemOpen
  \bibfield  {author} {\bibinfo {author} {\bibfnamefont {K.}~\bibnamefont
  {Kuroda}}, \bibinfo {author} {\bibfnamefont {M.}~\bibnamefont {Ye}}, \bibinfo
  {author} {\bibfnamefont {A.}~\bibnamefont {Kimura}}, \bibinfo {author}
  {\bibfnamefont {S.~V.}\ \bibnamefont {Eremeev}}, \bibinfo {author}
  {\bibfnamefont {E.~E.}\ \bibnamefont {Krasovskii}}, \bibinfo {author}
  {\bibfnamefont {E.~V.}\ \bibnamefont {Chulkov}}, \bibinfo {author}
  {\bibfnamefont {Y.}~\bibnamefont {Ueda}}, \bibinfo {author} {\bibfnamefont
  {K.}~\bibnamefont {Miyamoto}}, \bibinfo {author} {\bibfnamefont
  {T.}~\bibnamefont {Okuda}}, \bibinfo {author} {\bibfnamefont
  {K.}~\bibnamefont {Shimada}}, \bibinfo {author} {\bibfnamefont
  {H.}~\bibnamefont {Namatame}}, \ and\ \bibinfo {author} {\bibfnamefont
  {M.}~\bibnamefont {Taniguchi}},\ }\href {\doibase
  10.1103/PhysRevLett.105.146801} {\bibfield  {journal} {\bibinfo  {journal}
  {Phys. Rev. Lett.}\ }\textbf {\bibinfo {volume} {105}},\ \bibinfo {pages}
  {146801} (\bibinfo {year} {2010})}\BibitemShut {NoStop}%
\bibitem [{\citenamefont {Li}\ and\ \citenamefont {Carbotte}(2014)}]{ZLi:2014}%
  \BibitemOpen
  \bibfield  {author} {\bibinfo {author} {\bibfnamefont {Z.}~\bibnamefont
  {Li}}\ and\ \bibinfo {author} {\bibfnamefont {J.~P.}\ \bibnamefont
  {Carbotte}},\ }\href@noop {} {\bibfield  {journal} {\bibinfo  {journal}
  {Phys. Rev. B}\ }\textbf {\bibinfo {volume} {89}},\ \bibinfo {pages} {085413}
  (\bibinfo {year} {2014})}\BibitemShut {NoStop}%
\bibitem [{\citenamefont {Li}\ and\ \citenamefont {Carbotte}(2015)}]{ZLi:2015}%
  \BibitemOpen
  \bibfield  {author} {\bibinfo {author} {\bibfnamefont {Z.}~\bibnamefont
  {Li}}\ and\ \bibinfo {author} {\bibfnamefont {J.~P.}\ \bibnamefont
  {Carbotte}},\ }\href {\doibase 10.1103/PhysRevB.91.115421} {\bibfield
  {journal} {\bibinfo  {journal} {Phys. Rev. B}\ }\textbf {\bibinfo {volume}
  {91}},\ \bibinfo {pages} {115421} (\bibinfo {year} {2015})}\BibitemShut
  {NoStop}%
\bibitem [{\citenamefont {Bychkov}\ and\ \citenamefont
  {Rashba}(1984{\natexlab{a}})}]{Bychkov:1984}%
  \BibitemOpen
  \bibfield  {author} {\bibinfo {author} {\bibfnamefont {Y.~A.}\ \bibnamefont
  {Bychkov}}\ and\ \bibinfo {author} {\bibfnamefont {E.~I.}\ \bibnamefont
  {Rashba}},\ }\href@noop {} {\bibfield  {journal} {\bibinfo  {journal} {J.
  Phys. C: Solid State Phys.}\ }\textbf {\bibinfo {volume} {17}},\ \bibinfo
  {pages} {6039} (\bibinfo {year} {1984}{\natexlab{a}})}\BibitemShut {NoStop}%
\bibitem [{\citenamefont {Bychkov}\ and\ \citenamefont
  {Rashba}(1984{\natexlab{b}})}]{Bychkov:1984a}%
  \BibitemOpen
  \bibfield  {author} {\bibinfo {author} {\bibfnamefont {Y.~A.}\ \bibnamefont
  {Bychkov}}\ and\ \bibinfo {author} {\bibfnamefont {E.~I.}\ \bibnamefont
  {Rashba}},\ }\href@noop {} {\bibfield  {journal} {\bibinfo  {journal} {JETP
  Lett.}\ }\textbf {\bibinfo {volume} {39}},\ \bibinfo {pages} {78} (\bibinfo
  {year} {1984}{\natexlab{b}})}\BibitemShut {NoStop}%
\bibitem [{\citenamefont {Gusynin}\ \emph {et~al.}(2009)\citenamefont
  {Gusynin}, \citenamefont {Sharapov},\ and\ \citenamefont
  {Carbotte}}]{Gusynin:2009}%
  \BibitemOpen
  \bibfield  {author} {\bibinfo {author} {\bibfnamefont {V.~P.}\ \bibnamefont
  {Gusynin}}, \bibinfo {author} {\bibfnamefont {S.~G.}\ \bibnamefont
  {Sharapov}}, \ and\ \bibinfo {author} {\bibfnamefont {J.~P.}\ \bibnamefont
  {Carbotte}},\ }\href@noop {} {\bibfield  {journal} {\bibinfo  {journal} {New
  J. Phys.}\ }\textbf {\bibinfo {volume} {11}},\ \bibinfo {pages} {095013}
  (\bibinfo {year} {2009})}\BibitemShut {NoStop}%
\bibitem [{\citenamefont {Gusynin}\ \emph {et~al.}(2006)\citenamefont
  {Gusynin}, \citenamefont {Sharapov},\ and\ \citenamefont
  {Carbotte}}]{Gusynin:2006a}%
  \BibitemOpen
  \bibfield  {author} {\bibinfo {author} {\bibfnamefont {V.~P.}\ \bibnamefont
  {Gusynin}}, \bibinfo {author} {\bibfnamefont {S.~G.}\ \bibnamefont
  {Sharapov}}, \ and\ \bibinfo {author} {\bibfnamefont {J.~P.}\ \bibnamefont
  {Carbotte}},\ }\href@noop {} {\bibfield  {journal} {\bibinfo  {journal}
  {Phys. Rev. Lett.}\ }\textbf {\bibinfo {volume} {96}},\ \bibinfo {pages}
  {256802} (\bibinfo {year} {2006})}\BibitemShut {NoStop}%
\end{thebibliography}%

\end{document}